\documentclass[11pt]{article}
\usepackage[utf8]{inputenc}
\usepackage{graphicx}
\usepackage{amsfonts}
\usepackage{amsmath}
\usepackage{amssymb}
\usepackage{bm}
\usepackage{url}
\usepackage{fancyhdr}
\usepackage{indentfirst}
\usepackage{enumerate}
\usepackage{dsfont,footmisc}
\usepackage{comment}
\usepackage{subfigure}
\usepackage{caption}
\usepackage{multirow}
\usepackage{tabularx}

\usepackage{setspace}

\usepackage[misc]{ifsym}

\usepackage{amsthm}
\usepackage{color}
\usepackage{natbib}
\usepackage[british]{babel}
\usepackage[colorlinks=true,citecolor=blue]{hyperref}
\usepackage[right,pagewise,displaymath, mathlines]{lineno}

\def\dd{\mathrm{d}}

\newcommand{\var}{\mathrm{Var}}
\newcommand{\cov}{\mathrm{Cov}}

\newcommand{\E}{\mathbb{E}}
\newcommand{\R}{\mathbb{R}}

\newcommand{\p}{\mathbb{P}}

\renewcommand{\(}{\left(}
\renewcommand{\)}{\right)}

\newcommand{\esssup}{\mathrm{ess\mbox{-}sup}}
\newcommand{\essinf}{\mathrm{ess\mbox{-}inf}}
\renewcommand{\ge}{\geqslant}
\renewcommand{\le}{\leqslant}
\renewcommand{\geq}{\geqslant}
\renewcommand{\leq}{\leqslant}
\renewcommand{\epsilon}{\varepsilon}

\theoremstyle{plain}
\newtheorem{theorem}{Theorem}[section]
\newtheorem{corollary}{Corollary}[section]
\newtheorem{lemma}{Lemma}[section]
\newtheorem{proposition}{Proposition}[section]

\theoremstyle{definition}
\newtheorem{definition}{Definition}[section]

\theoremstyle{remark}
\newtheorem{remark}{Remark}[section]

\theoremstyle{definition}

\numberwithin{equation}{section} \numberwithin{theorem}{section}
\renewcommand{\cite}{\citet}

\DeclareMathOperator*{\argmin}{arg\,min}
\title{Worst-Case Values of  Target Semi-Variances   With   Applications  to Robust Portfolio Selection}
\author{Jun Cai\thanks{Department of Statistics and Actuarial Science, University of Waterloo,  Canada. \Letter~{\url{jcai@uwaterloo.ca}}}, \  \  \ \  Zhanyi Jiao\thanks{Department of Mathematics,
Illinois State University, Normal, Illinois, USA. \Letter~{\url{zjiao1@ilstu.edu}}},  \ \ \  \
Tiantian Mao\thanks{Department of Statistics and Finance, School of Management,
University of Science and Technology of China, China. \Letter~{\url{tmao@ustc.edu.cn}}}}
\date{\today}

\begin{document}

\maketitle

\vspace{-0.5 cm} 
	\begin{abstract}
	
The expected regret and target semi-variance  are two of the most important risk measures for downside risk. When the distribution of a loss is uncertain, and only partial information of the loss is known, their worst-case values play important roles in robust risk management for finance, insurance,  and many other fields.   \cite{J77} derived the   worst-case   expected regrets 
      when    only the mean and variance of a loss are known  and the  loss is arbitrary,   
          symmetric,  or non-negative. While \cite{CHZ11} obtained   
   the  worst-case  target semi-variances under similar conditions but focusing on 
      arbitrary losses.   
      In this paper, we first complement the study  of  \cite{CHZ11}  on  the  worst-case  target semi-variances  and derive the closed-form expressions  for  the worst-case  target semi-variance    when   
    only the mean and variance of a loss are known and 
     the loss is      symmetric or non-negative. Then, we investigate  worst-case target semi-variances over uncertainty sets that represent  
      undesirable scenarios faced by an investors.        Our methods 
    for deriving these   worst-case values  are different from those used in  \cite{J77} and  \cite{CHZ11}.        
   As applications of the results derived in this paper, we propose  robust portfolio selection methods that minimize the worst-case target semi-variance of a portfolio loss over different uncertainty sets.
   To explore the insights of our robust portfolio selection methods, we conduct numerical experiments with real financial  data and compare our portfolio selection methods with several existing  portfolio selection models related to the models proposed in this paper.   \\

  	\end{abstract}

\newpage

\section{Introduction}
\label{sec:tsv-intro}

Assume that  $X$ is  a  random variable  denoting   the loss of an investment  portfolio. Hence, in this 
paper,    positive values of $X$ represent losses and negative values of $X$ represent gains or returns. The manager of an  investment portfolio  often has a   target return $-t$ or equivalently  a   threshold loss  $t$. Thus,  the loss function  $(X-t)_+$  represents  the 
     downside risk/loss of the    portfolio, while  $(-X-(-t))_+ =  (X-t)_-$  denotes   the 
    excess profit of  the    portfolio over the target return.   Here and throughout this paper,  
    $(x)_+=\max\{x, 0\}$  and  $(x)_-=\max\{-x, 0\}$ for any $x \in \R=(-\infty, \infty)$.   
    Two  important quantities of the downside risk $(X-t)_+$  are the first moment $\E[(X-t)_+]$   that measures the expected  loss above 
    the threshold loss   $t$  and  the second  
    moment  $\E[(X-t)_+^2]$  that quantifies the dispersion of the loss  that exceeds   the threshold loss    $t$. 
    In the literature,  the two moments are often called the first-order  and second-order upper partial moments, respectively.  
   In addition,    the first  moment  $\E[(X-t)_+]$ is also referred to as the {\it expected  regret}   or  {\it target shortfall} (see, e.g., 
   \cite{TU04},     \cite{K11}),   while 
    the second moment  $\E[(X-t)_+^2]$   is also referred to as    the {\it target semi-variance} (see, e.g., \cite{RVK11}). If the target return  is equal to  
       the expected return, namely,  $t=\E[X]$,     the target semi-variance    $\E[(X-\E[X])_+^2]$ is called the semi-variance of the loss $X$.
 Both of the expected regret   and   the target semi-variance      are important 
risk measures of the downside risk and have been  extensively  used  in finance,  insurance, operations research, 
and  many other fields. 

If  the `true' distribution of  the loss $X$ is known, the expected  regret  $\E[(X-t)_+]$ and   the target semi-variance    $\E[(X-t)_+^2]$ 
can be calculated  analytically or numerically. However, in practice, the `true' distribution of $X$ is often  unknown. A decision maker 
may have only partial information on $X$ such as the mean and  variance of $X$. If  only partial information on $X$ 
 is available and the possible distributions of $X$ belong to a   
 distribution set  ${\cal L}$,  called  an {\it uncertainty set}  for $X$,  
 a decision maker is  often interested in   
 $\sup_{F \in {\cal L}}\E^F[(X-t)_+]$ and   $\sup_{F \in {\cal L}}\E^F[(X-t)^2_+]$,    which 
 are respectively called the {\it worst-case  expected regret}     and  
  the {\it worst-case  target semi-variance}   
  over the uncertainty  set  ${\cal L}$. Here and throughout this paper,  for a function $h$ defined on $\R$ and a risk measure $\rho$,  such as 
  expectation $\E$, variance $\rm Var$, and  conditional value-at-risk CVaR,  $\rho^{F}[h(X)]$ means that the risk measure  of 
$\rho(h(X))$   is calculated under the distribution $F$ if the distribution of  $X$ is $F$.  
      In the literature,   for a random variable $X$ and a loss/cost function $h$, when the `true' distribution of $X$  is unknown or uncertain 
 but is assumed to be in  an uncertainty set ${\cal L}$, the  optimization problem of  
    \begin{equation}
 \label{FS-fX}
 \sup_{F \in {\cal L}} \rho^F[h(X)] 
 \end{equation}
is called 
 a  distributionally  robust optimization (DRO) problem,  and   if there exists a distribution $F^* \in {\cal L}$ such 
 that  $\sup_{F \in {\cal L}} \rho^F[h(X)] =  \rho^{F^*}[h(X)]$, such a distribution is called a {\it worst-case 
 distribution}.   The DRO problem \eqref{FS-fX}  and its applications 
  have  been extensively  studied in the literature of finance, insurance, operations research,  and many other fields.  For instance,  \cite{J77} 
  investigated   problem \eqref{FS-fX} when  $\rho=\E$,  $h(x)=(x-t)_+$,  
  ${\cal L}$ is a  set containing  distributions with the given first two  moments,   and 
  $X$   is an arbitrary,    
  symmetric or non-negative random variable.   
  \cite{ZPD09}      considered  problem \eqref{FS-fX} when $\rho=\E$, 
   $h(x)=(x-t)_+$,  ${\cal L}$ is a  set
 containing  distributions with the given first three moments. 
 \cite{CHZ11}  studied  problem \eqref{FS-fX} when  $\rho=\E$,    $h(x)=(x-t)_-^2$,   and    ${\cal L}$ is a set 
 containing  distributions with the given first two moments.   \cite{TY23}     discussed  problem \eqref{FS-fX} when   $h(x)=x^m$ or $(x-t)^m_+$,  $m=1,2,...,$  
 and ${\cal L}$ is a set containing 
  distributions satisfying  a distance constraint to  a reference distribution.  \cite{CLY24}   studied problem \eqref{FS-fX} when  $\rho$ is a 
 distortion risk measure, $h(x)=(x-t)_+$, and  ${\cal L}$ is a set 
 containing  distributions    satisfying  a distance constraint to  a reference distribution and 
   constraints on first two  moments. For  the   studies  and applications of  the DRO problem \eqref{FS-fX}  with   other forms of the function $h$ and the risk measure $\rho$, we  refer to \cite{BN98},   \cite{BP02},  \cite{GOO03},   
    \cite{H05}, 
     \cite{NPS08},   \cite{ZF09}, \cite{ZLW09},   \cite{A17},  
       \cite{P17},    
\cite{Li2018}, \cite{KLLZ19},   \cite{LM22},  \cite{B23},    \cite{CLM23}, and the references therein. 

\begin{figure}[t]
\centering
\subfigure{\includegraphics[width=6cm]{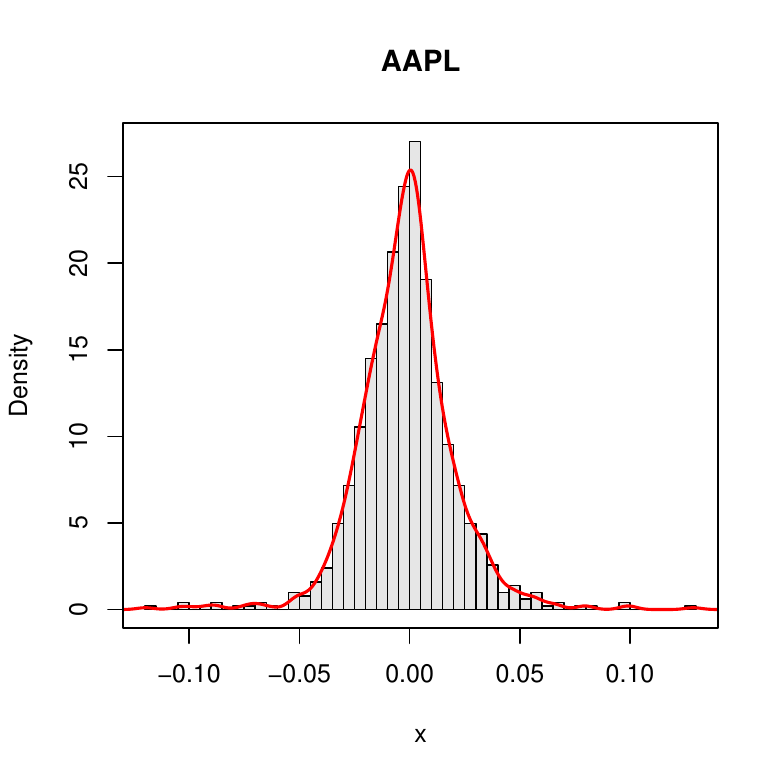}}
\subfigure{\includegraphics[width=6cm]{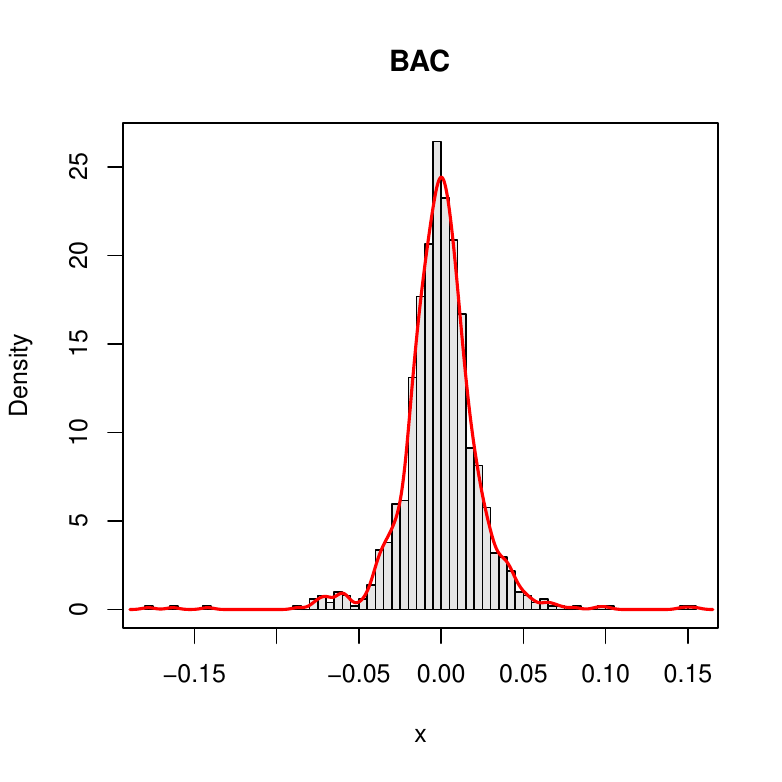}}
\subfigure{\includegraphics[width=6cm]{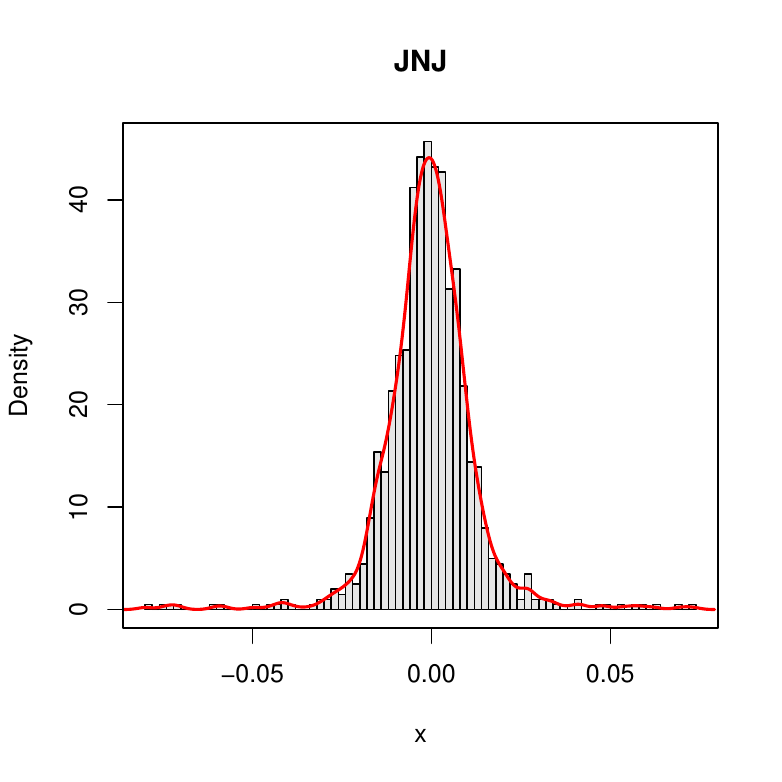}}
\subfigure{\includegraphics[width=6cm]{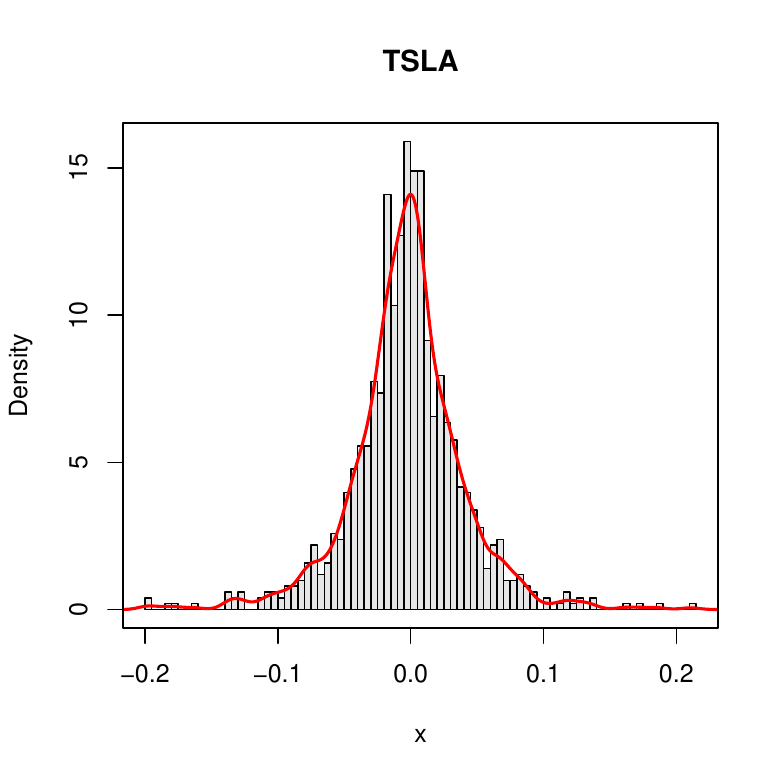}}

\caption{Histograms of daily losses of  the  stocks of  Apple (AAPL), Bank of America (BAC), Johnson \& Johnson (JNJ) and Tesla (TSLA). The data used for this figure covers a four-year period from January 2, 2019, to January 2, 2023, and includes 1007 observations of daily losses from Yahoo! Finance.} \label{fig:sym}
    \end{figure}

 In many DRO problems, it is  assumed that the mean and variance or second moment of a random variable $X$ are the 
 only known 
 information on the distribution of $X$, which correspond  to      the following uncertainty set 
  \begin{align}
 \label{L}
     \mathcal{L}(\mu, \sigma) &= \big \{ F \in \mathcal{F}(\R): \, \int^{\infty}_{-\infty} x \, \dd F(x) = \mu, \int^{\infty}_{-\infty} x^2 \, \dd F(x) = \mu^2 + \sigma^2 \big \} \nonumber \\
     &= \big \{ F \in \mathcal{F}(\R): \, \E^F[X]=\mu, \, \E^F[X^2] = \mu^2 + \sigma^2 \big \},
     \end{align}
  where $ \mathcal{F}(\R)$ is the set of all the distributions defined on 
     $\R$. 
 In  practice, a decision maker may have additional  information on  the distribution of $X$  besides  its mean and variance. In finance, a decision maker may notice that the loss  data have the symmetric  features. For instance, Figure \ref{fig:sym} displays the histograms of  daily losses of  the stocks of  Apple, Bank of America, Johnson \& Johnson, and Tesla. The daily losses  of these stocks exhibit a high degree of symmetry. 

In fact, in many portfolio selection researches, the daily   losses of the underlying assets  are assumed to have multivariate symmetric distributions such as multivariate normal distributions, multivariate $t$-distributions, multivariate elliptical distributions, and so on. See,  for example, 
   \cite{OR83}, \cite{B08}, \cite{HK10}, \cite{F18}, and the references therein.

In addition,  in insurance, loss random variables often are  the  amounts and numbers of insurance claims that are 
 non-negative random variables. 
Hence,  the following two uncertainty sets 
    \begin{align}
    &\mathcal{L}_{S}(\mu, \sigma) = \big \{F \in \mathcal{L}(\mu, \sigma):  \ 
          \text{$F$ is symmetric} \label{LS} \big \}, \\
    &\mathcal{L}^{+}(\mu, \sigma) =\big \{ F \in \mathcal{L}(\mu, \sigma):  
         \  F(0-)=0 \big \}, \label{L+}
\end{align}
are also interesting in the  study of DRO problems. In this paper,  the formal  definitions  of symmetric distributions  are given in Definitions     \ref{sdf-u} and \ref{sdf-m}, 
and  a  non-negative distribution  means that $F(0-)=\p\{X <0\}=0$ or $F$ is a distribution of 
a non-negative random variable $X$.

The   closed-form expressions  for  $\sup_{F \in {\cal L}} \E[(X-t)_+]$  have been derived in \cite{J77} when 
${\cal L}$ is one of  the three uncertainty sets  $\mathcal{L}(\mu, \sigma)$  
$\mathcal{L}_{S}(\mu, \sigma)$, and  $\mathcal{L}^{+}(\mu, \sigma)$. 
The  closed-form expression for  $\sup_{F \in {\cal L}} \E[(X-t)^2_-]$  has been  obtained in      \cite{CHZ11} when 
${\cal L} = \mathcal{L}(\mu, \sigma)$. To the best of  our knowledge,    the worst-case values  of $\E[(X-t)^2_+]$ over the uncertainty sets  
$\mathcal{L}_{S}(\mu, \sigma)$ and  $\mathcal{L}^{+}(\mu, \sigma)$    have not been solved. 
As discussed later in this paper,   the methods and proofs used in   \cite{J77}  and \cite{CHZ11}    do not apply for the worst-case values  of $\E[(X-t)^2_+]$ over the uncertainty sets  $\mathcal{L}_{S}(\mu, \sigma)$ and  $\mathcal{L}^{+}(\mu, \sigma)$.

In this paper, first, we  complement 
the study of \cite{CHZ11} on worst-case values of 
the  target semi-variance   and obtain the   closed-form expressions for the  worst-case values of 
the  target semi-variance over the  uncertainty sets  $\mathcal{L}_{S}(\mu, \sigma)$ and  $\mathcal{L}^{+}(\mu, \sigma)$.
Second, motivated by  the classical mean-variance (M-V)  portfolio selection model, we discuss the applications of the worst-case target semi-variance in portfolio selection problems and propose portfolio selection models based on expected excess profit-target semi-variance (EEP-TSV). These models aim to minimize the worst-case target semi-variance of portfolio loss over undesirable scenarios where the expected excess profit does not meet a desirable minimum level  $\lambda$. 
As illustrated  in Section \ref{sec:ER-TSV}, 
   these  EEP-TSV-based portfolio selection models can be  transformed  into  
   problems that minimize the  worst-case   target semi-variance of  portfolio loss over the   
   following uncertain sets: 
      \begin{align}
    \mathcal{L}_{\lambda}(\mu, \sigma) 
  &= \big \{  F \in \mathcal{L}(\mu, \sigma):   \    
  \E^F[(X-t)_-]
  \leq \lambda  \big \},    \label{eq:budget<lambda} \\
  \mathcal{L}_{S, \lambda}(\mu, \sigma) &= \big \{ F \in \mathcal{L}(\mu, \sigma):   \   \text{$F$ is symmetric and} \      \E^F[(X-t)_-]
  \leq \lambda   \},  \label{LS<lambda}  \\
  \mathcal{L}^{+}_{\lambda}(\mu, \sigma) & =\big \{ F \in \mathcal{L}(\mu, \sigma):  
         \  F(0-)=0  \ \text{and} \      \E^F[(X-t)_-]
  \leq \lambda   \big \}. \label{L+<lambda}   
           \end{align}    
  Note that, in this paper,  $\E^{F}[(X-t)_-] $ denotes the   expected  excess 
 profit. Thus,  if $\lambda > 0$  represents  a desirable minimum  level for  the expected  excess 
 profit, 
         then  $\E^{F}[(X-t)_-] \le \lambda$  in   \eqref{eq:budget<lambda}-\eqref{L+<lambda} indicates 
          that  the expected  excess 
 profit      does not  meet  the  desirable minimum  level, which is an undesirable scenario for an investor.  
 Conversely,     $\E^{F}[(X-t)_-]  >  \lambda$  represents  a desirable scenario for an investor.
   In fact,  for any uncertainty set $\mathcal{L}$ for the loss random variable $X$, it holds that 
  \[
    \mathcal{L} 
  = \big \{  F \in \mathcal{L}:   \    
  \E^F[(X-t)_-]
  \leq \lambda  \big \}  \, \cup  \, \big \{  F \in \mathcal{L}:   \    
  \E^F[(X-t)_-]
  >  \lambda  \big \}.   
 \]
An investor   is primarily 
 concerned with   the undesirable scenario where   $ \E^F[(X-t)_-]
  \leq \lambda $  and  with  the worst-case values  over the set $ \big \{  F \in \mathcal{L}:   \    
  \E^F[(X-t)_-]
  \leq \lambda  \big \} $ 
   such as those   defined in  \eqref{eq:budget<lambda}-\eqref{L+<lambda}.
 In addition, the uncertainty sets   $\mathcal{L}(\mu, \sigma)$,   
   $\mathcal{L}_{S}(\mu, \sigma),$ and   $\mathcal{L}^+(\mu, \sigma)$  can be treated as   the limiting cases of  
   $\mathcal{L}_{\lambda}(\mu, \sigma)$, 
   $ \mathcal{L}_{S, \lambda}(\mu, \sigma)$, and  
   $\mathcal{L}^+_{ \lambda}(\mu, \sigma)$,    respectively, as $\lambda  \to \infty$.

The uncertainty sets $\mathcal{L}_{S}(\mu, \sigma)$, $\mathcal{L}^+(\mu, \sigma)$,   $\mathcal{L}_{\lambda}(\mu, \sigma)$, and  $\mathcal{L}_{S,\lambda}(\mu, \sigma)$ have more constraints than $\mathcal{L}(\mu, \sigma)$.  Finding  the worst-case values of  $\E[(X-t)^2_+]$ over the uncertainty  sets  $\mathcal{L}_{S}(\mu, \sigma)$, $\mathcal{L}^+(\mu, \sigma)$,  $\mathcal{L}_{\lambda}(\mu, \sigma)$,  $\mathcal{L}_{S, \lambda}(\mu, \sigma)$  is 
    a  challenging question, in particular, over the uncertainty  sets  $\mathcal{L}_{S}(\mu, \sigma)$ and   $\mathcal{L}_{S,\lambda}(\mu, \sigma)$.    The main method   used in this paper for    finding  these  worst-case values  is to  reformulate  these  infinite-dimensional  optimization problems to  finite-dimensional  optimization problems and then solve the finite-dimensional  optimization problems to obtain the  closed-form expressions for the worst-case values.

The rest of paper is structured as follows. In Section \ref{sec-pm}, we give the preliminaries of the worst-case values of 
the expected regret and target semi-variance and  describe our motivation      
  for studying the worst-case target semi-variance.  In Section \ref{WC-SD},     we derive the  closed-form  expressions for  the  worst-case   target semi-variance 
over the uncertainty set $\mathcal{L}_{S}(\mu, \sigma)$. In Section \ref{sec:WC-ER}, 
 we provide  the closed-form  expressions for  worst-case target semi-variances  over
   undesirable scenarios, which are  worst-case target semi-variances over the uncertainty sets 
  $\mathcal{L}_{\lambda}(\mu, \sigma)$ and $\mathcal{L}_{S, \lambda}(\mu, \sigma)$. 
  In Section \ref{sec:ER-TSV}, we propose robust portfolio selection models that minimize the target semi-variance under the different uncertainty sets discussed above.
     In Section \ref{sec:num},  we use the real finance data to compare the investment performances of our portfolio selection methods with several 
   existing portfolio selection models related to the models proposed in this paper.  In Section \ref{sec:end}, we give concluding remarks.  
   The proofs of the main results in this paper  are  presented in the appendix, which is located in Section \ref{sec:App}.

\section{Preliminary and motivation} 
\label{sec-pm}
   \begin{definition}
\label{sdf-u}
The distribution $F$ of a random variable $X$ is said  to  be symmetric if there exists a constant
  $a$ such that   $ \mathbb{P}
(X - a> x) = \mathbb{P}(X-a < -x)$,  under the distribution $F$,  for all $x \in  \mathbb{R}$. If such a constant $a$ exists,
 random variable $X$ or its distribution  is said to be symmetric about   $a$.  \hfill $\Box$
    \end{definition}
    
Intuitively, random variable $X$ is  symmetric about   $a$ if and only if  $X-a$   is  symmetric about  the origin of $\R$. Examples of  continuous symmetric distributions include the 
 Cauchy distribution, normal distribution, $t$-distribution, uniform distribution,  logistic distribution,  and many others. 
Examples of  discrete  symmetric distributions include discrete uniform distribution, $k$-point symmetric distribution (where $k \ge 2$ is an integer),  and many others. In addition, a degenerate distribution is also symmetric according to Definition \ref{sdf-u}.

To give a detailed review of the known results about  the worst-case values $\sup_{F \in {\cal L}} \E^F[(X-t)_+]$ and  $\sup_{F \in {\cal L}} \E^F[(X-t)^2_+]$ and illustrate our motivation for studying the worst-case target semi-variance,    we  state the  results of  \cite{J77} and \cite{CHZ11}   about these   worst-case values  and give remarks on these results and their proofs below. 
    \begin{proposition} (\cite{J77}) 
\label{lem:first}
For any $\mu, t \in \R$ and $\sigma \in \R^+=(0, \infty)$,   if the uncertainty set of  random variable $X$ is $\mathcal{L}(\mu, \sigma)$, then
\begin{equation}
\label{upm1}    \sup_{F \in \mathcal{L}(\mu, \sigma)} \E^{F}[(X-t)_{+}]= \frac{1}{2} \left( \mu - t + \sqrt{\sigma^2 + (\mu - t)^2}\right).
\end{equation}
If the uncertainty set of $X$ is $\mathcal{L}_S(\mu, \sigma)$, then
\begin{equation}
\label{upm1S}
    \sup_{F \in \mathcal{L}_S(\mu, \sigma)} \E^{F}[(X-t)_{+}] =     \begin{cases}
    \frac{8(\mu -t)^2 + \sigma^2}{8(\mu -t)}, & t < \mu - \frac{\sigma}{2}, \\
    \frac{1}{2}(\mu + \sigma -t), & \mu - \frac{\sigma}{2} \leq t < \mu + \frac{\sigma}{2},\\
    \frac{\sigma^2}{8(t-\mu)}, & t \geq \mu + \frac{\sigma}{2}.
       \end{cases}
\end{equation}
If the uncertainty set of $X$  is $\mathcal{L}^+(\mu, \sigma)$ and $\mu >0$,
then
\begin{equation}
\label{upm1+}
    \sup_{F \in \mathcal{L}^+(\mu, \sigma)}  \E^{F}[(X-t)_{+}] =
    \begin{cases}
     \mu - t, & t < 0, \\
  \mu - \frac{\mu^2 t}{\sigma^2+\mu^2}, & 0 \leq t < \frac{\sigma^2+\mu^2}{2\mu},\\
     \frac{1}{2}\left(\mu - t + \sqrt{\sigma^2 + (\mu - t)^2}\right), & t \geq \frac{\sigma^2+\mu^2}{2\mu}.
       \end{cases}
\end{equation}
\end{proposition}

\begin{remark} The main idea of  \cite{J77}'s 
   proof for Proposition \ref{lem:first} as follows:   First   apply  Cauchy-Schwarz's inequality for $\E[(X-t)_+]$  or start  with  
   $(\E[(X-t)_{+}])^2 = \big ( \int^{\infty}_{t} (x-t) \, \dd F(x) \big )^2 \leq \int^{\infty}_{t} \,   \dd F(x) \,  \int^{\infty}_{t} (x-t)^2 \,  \dd F(x) $ 
and  obtain the sharp  upper bound for $\sup_{F \in {\cal L}} \int^{\infty}_{t} \,   \dd F(x) \,  \int^{\infty}_{t} (x-t)^2 \,  \dd F(x)$, 
 and then  verify  that  the   upper bound is also the sharp bound for   $\sup_{F \in {\cal L}} \E^F[(X-t)_+]$. We point out that  
  the arguments and proofs used in \cite{J77} for Proposition \ref{lem:first} do not work  for   the worst-case values of 
$\sup_{F \in {\cal L}} \E^F[(X-t)^2_+]$ when ${\cal L}$ is any of the uncertainty sets  $\mathcal{L}(\mu, \sigma)$,
  $\mathcal{L}_S(\mu, \sigma)$,  and $\mathcal{L}^+(\mu, \sigma)$. If fact,  
   if one applies    Cauchy-Schwarz's inequality  to $\E[(X-t)^2_{+}]$, one obtains 
   $(\E[(X-t)^2_{+}])^2 = \big ( \int^{\infty}_{t} (x-t)^2 \, \dd F(x) \big )^2 \leq \int^{\infty}_{t} \,  \dd F(x) \, 
   \int^{\infty}_{t} (x-t)^4 \,  \dd F(x).$ 
However, the      upper bound     $\int^{\infty}_{t} \,  \dd F(x) \, 
   \int^{\infty}_{t} (x-t)^4 \,  \dd F(x) $  does not provide any useful information for 
   $ (\E[(X-t)^2_{+}])^2$ since  
 $\int^{\infty}_{t} (x-t)^4 \,  \dd F(x) $  may be equal to $\infty$ when  $F$ is in any of the  sets  $\mathcal{L}(\mu, \sigma)$,
  $\mathcal{L}_S(\mu, \sigma)$,  and $\mathcal{L}^+(\mu, \sigma)$. \hfill $\Box$ 
   \end{remark} 
 
Note that for any $x \in \R$,
$(x)_+ -
(x)_- = x$, $(x)^2_+ +
(x)_-^2 = x^2$,  and $(x)_+ =(-x)_-$. Hence, for any $t \in \R$, 
     if   $ \mathcal{L}^*(\mu, \sigma) \subset {\cal F}(\R)$ is 
 a set of distributions satisfying that for any $F \in \mathcal{L}^*(\mu, \sigma)$, $\E^F[X]=\mu$ and  
 $\E^F[X^2]=\mu^2+\sigma^2$, then  
 \begin{align}
  \sup_{F \in \mathcal{L}^*(\mu, \sigma)}   \E^F[(X-t)_{+}] & \ = \     \sup_{F \in \mathcal{L}^*(\mu, \sigma)} \E^F[(X-t)_-]   + \mu-t, \label{sup-sup-1st}
   \end{align}
   \begin{align}
 \inf_{F \in \mathcal{L}^*(\mu, \sigma) } \E^{F}[(X-t)_{+}]
 & \ = \        \inf_{F \in \mathcal{L}^*(\mu, \sigma)} \E^{F}[(X-t)_-]   + \mu-t,     \label{inf-inf-1st} \\
  \sup_{F \in \mathcal{L}^*(\mu, \sigma)} \E^{F}[(X-t)_{+}^2]
   & \ = \ \sigma^2 +  (t-\mu)^2 \, -  \, \inf_{F \in \mathcal{L}^*(\mu, \sigma)} \E^{F}[(X-t)^2_-], \label{sup-inf-2nd} \\
  \inf_{F \in \mathcal{L}^*(\mu, \sigma)} \E^{F}[(X-t)_{+}^2]
   & \ = \ \sigma^2 +  (t-\mu)^2 \, - \,   \sup_{F \in \mathcal{L}^*(\mu, \sigma)} \E^{F}[(X-t)^2_-]. \label{inf-sup-2nd}
 \end{align}
In addition,    for random variable $X$,  the conditions of $\E[X]=\mu$ and $\E[X^2] =\mu^2+\sigma^2$ are equivalent to the conditions of $\E[-X]=-\mu$ and $\E[(-X)^2] =\mu^2+\sigma^2$, and the condition that $X$ is symmetric about $\mu$ is  equivalent to the  condition that  $-X$ is symmetric about $-\mu$. Hence, 
 if   $\mathcal{L}_0(\mu, \sigma)$ is one of the sets
$ \mathcal{L}(\mu, \sigma)$ and   $\mathcal{L}_{S}(\mu, \sigma)$,  
then for $k=1, 2$, 
 \begin{align}
\sup_{F \in \mathcal{L}_0(\mu, \sigma) } \E^F[(X-t)^k_+] &
   =\sup_{F \in \mathcal{L}_0(-\mu, \sigma) } \E^F[(X-(-t))^k_-)], \label{sup-h-X-t} \\
    \inf_{F \in \mathcal{L}_0(\mu, \sigma) } \E^F[(X-t)^k_+)] &    
    =\inf_{F \in \mathcal{L}_0(-\mu, \sigma) } \E^F[(X-(-t))^k_-)].  \label{inf-h-X-t}
  \end{align}
   We also point out that the downside risk of an investment portfolio  is $(t-X)_+ = (X-t)_-$ if $X$ 
 represents the return or gain of the portfolio and $t$ is the target return.

\begin{proposition} (\cite{CHZ11})
\label{lem:second}
For any $\mu, t \in \R$ and $\sigma \in \R^+$,  we have 
     \begin{align} \sup_{F \in \mathcal{L}(\mu, \sigma)} \E^{F}[(X-t)^2_-]  &= \sigma^2 + (t-\mu)^2_+,     \label{(x-t)_-^2} \\ 
\sup_{F \in \mathcal{L}(\mu, \sigma)} \E^{F}[(X-t)_{+}^2]
    &
    = \sigma^2 + 
     (\mu-t)_+^2.   \label{sup-1st}
    \end{align}
       \end{proposition}
  \begin{remark} Formula \eqref{(x-t)_-^2} was proved  by \cite{CHZ11}  by the following idea: For 
  $F \in \mathcal{L}(\mu, \sigma)$,  (i) 
  yield  a lower bound for   
   $\E[(X-t)^2_+]$ by  using Jensen's inequality   and  then obtain an 
    upper bound for $ \E[(X-t)^2_{-}]$ by using  the equation  $ \E[(X-t)^2_{-}] = 
\E[(X-t)^2] - \E[(X-t)^2_+] = \sigma^2 + (\mu- t)^2 -  \E[(X-t)^2_+]$, and (ii) verify the  upper bound for $ \E[(X-t)^2_{-}]$ 
is sharp for  $\sup_{F \in \mathcal{L}(\mu, \sigma)} \E^{F}[(X-t)^2_-] $.    
Formula  \eqref{sup-1st}  follows directly  by applying   \eqref{sup-h-X-t} to \eqref{(x-t)_-^2}.  Following 
the proof of  \cite{CHZ11} for \eqref{(x-t)_-^2},  we  can show  that   
$ \sup_{F \in \mathcal{L}(\mu, \sigma)} \E^{F}[(X-t)_{+}^2] = \sup_{F \in \mathcal{L}^+(\mu, \sigma)} \E^{F}[(X-t)^2_-]$ that is proved in Corollary  \ref{coro:positive} in this paper.    However, the  proofs used in 
  \cite{CHZ11} for the worst-case value $ \sup_{F \in \mathcal{L}(\mu, \sigma)} \E^{F}[(X-t)^2_-]$ do not work for the worst-case values $ \sup_{F \in \mathcal{L}_S(\mu, \sigma)} \E^{F}[(X-t)^2_+]$   since  Jensen's inequality cannot yield  
   tight upper bound for    $ \sup_{F \in \mathcal{L}_S(\mu, \sigma)} \E^{F}[(X-t)^2_+]$.  
In Theorem   \ref{th-2nd-TVS-S} of this paper, we derive the explicit  and closed-form expression for $ \sup_{F \in \mathcal{L}_S(\mu, \sigma)} \E^{F}[(X-t)^2_+]$ by a method different from those used in   \cite{J77} and \cite{CHZ11}.    
      \hfill $\Box$
       \end{remark}

\section{Worst-case  target semi-variances with symmetric distributions } 
\label{WC-SD}

In this section, we solve the worst-case  target semi-variance over the  uncertainty set \eqref{LS} 
with  symmetric distributions, which is the following optimization problem: 
  \begin{equation}
\begin{aligned}
\label{pro:tsv}
    &\sup_{F \in \mathcal{L}_{S}(\mu, \sigma)}  \,\,
     \E^{F}[(X-t)_{+}^2]. 
 \end{aligned}
\end{equation}
  Problem   \eqref{pro:tsv} is an infinite-dimensional optimization problem. However, we are able  to 
    reformulate it
   as a finite-dimensional optimization problem and solve the
    finite-dimensional optimization problem.  To do so, we introduce the definition and notation for   a
$k$-point (discrete) distribution.  
 \begin{definition} 
 \label{de-k-ds}
Let $
 [x_1,p_1;\ldots; x_k,p_k]
$
denote the probability function of   a $k$-point (discrete)  random variable $X$,  where   $\p(X=x_i)=p_i$, $0 \le p_i <1$,   $i=1,\ldots,k$,  $\sum_{i=1}^k p_i=1$,  $k \ge 2$.
\hfill $\Box$
 \end{definition} 
We point out that according to Definition  \ref{de-k-ds},  when   $k \ge 3$,  a $k$-point distribution may also be   a
$l$-point  distribution, where   $2 \le l \le  k$.  
   \begin{lemma}
\label{lem:3}
For any $F\in \mathcal{L} (\mu, \sigma)$ with $\sigma>0$, there exists a two-point   distribution  $F^*\in \mathcal{L}(\mu, \sigma)$ such that   the support of $F^*$ belongs to $[\essinf  \, F,\, \esssup  \, F]$. Moreover, if $F$ is symmetric, then such  a two-point   distribution  $F^*$ can be chosen as a symmetric one.
\end{lemma}
 
   Lemma \ref{lem:3} implies   that  for any distribution $F$  with given mean and variance, there exists 
 a two-point   distribution $F^*$ such that  $F^*$ has the same mean and variance as $F$ and the support of 
 $F^*$  is  contained within    the support of $F$. This  lemma   plays a key role in solving  problem \eqref{pro:tsv} or proving the following  Theorem \ref{th-2nd-TVS-S}, which is the main result of this section.  The proof of Lemma \ref{lem:3}  is given 
 in the appendix.

    \begin{remark} 
   \label{re-mu=0}
 Note that  $\E[X] =\mu$ and  $\var(X) =\sigma^2$ if and only if  $\E[X-\mu]=0$,  $\var(X-\mu) =\sigma^2$, and  $X$ with mean $\mu$  is   a symmetric distribution   if and only if $X-\mu$ is symmetric about 0. 
 Hence,
   if
$
 \sup_{F \in \mathcal{L}^*(0, \sigma)} \E^{F}[h(X-t)]=  U(t,\sigma)$ and 
 $\inf_{F \in \mathcal{L}^*(0, \sigma)} \E^{F}[h(X-t)]   =  L(t,\sigma),
$
 then
 \begin{align}
 \label{sup-inf-h-U-L}
  \sup_{F \in \mathcal{L}^*(\mu, \sigma)} \E^{F}[h(X-t)]   &= U(t-\mu, \sigma), \ \ \ \ \ \   \inf_{F \in \mathcal{L}^*(\mu, \sigma)} \E^{F}[h(X-t)]   
  = L(t-\mu, \sigma),    
   \end{align}
  where   $\mathcal{L}^*(\mu, \sigma)$ is one of the sets
$ \mathcal{L}(\mu, \sigma)$,  $\mathcal{L}_{S}(\mu, \sigma)$, and 
$ \mathcal{L}_{\lambda}(\mu, \sigma)$;     $h$ is a function defined on $\R$;  and $t \in \R$.  
Thus,  without loss of generality, we  can assume $\mu=0$ in 
$ \mathcal{L}(\mu, \sigma)$,  $\mathcal{L}_{S}(\mu, \sigma)$, and $ \mathcal{L}_{\lambda}(\mu, \sigma)$ 
 when solving the optimization  problems   $\sup_{F \in \mathcal{L}^*(\mu, \sigma)} \E^{F}[h(X-t)] $ and   
$\inf_{F \in \mathcal{L}^*(\mu, \sigma)} \E^{F}[h(X-t)]$. \hfill $\Box$
\end{remark}

\begin{theorem}
\label{th-2nd-TVS-S}
For $\mu, t \in \R$ and $\sigma \in \R^{+}$, we have
   \begin{equation}
   \label{upm2S}
    \sup_{F \in \mathcal{L}_{S}(\mu, \sigma)}\E^{F}[(X-t)_{+}^2]=
        \begin{cases}
    \sigma^2 + (t-\mu)^2, & t \le  \mu-\sigma,\\
    \frac{1}{2}(\mu- t+\sigma)^2, & \mu-\sigma <  t\leq \mu,\\
    \frac{\sigma^2}{2}, & t > \mu.
    \end{cases}
\end{equation}
    \end{theorem}

The main idea for the proof of Theorem \ref{th-2nd-TVS-S} is to use Lemma \ref{lem:3} and   reformulate the problem  
$ \sup_{F \in \mathcal{L}_{S}(\mu, \sigma)}\E^{F}[(X-t)_{+}^2]$  as the problem  $ \sup_{F \in \mathcal{L}_{k, S}(\mu, \sigma)}\E^{F}[(X-t)_{+}^2] $, where 
the set $\mathcal{L}_{k, S}(\mu, \sigma) = \{F \in \mathcal{L}_{S}(\mu, \sigma): \, F \ \mbox{is a $k$-point distribution}\} $ is a subset of $\mathcal{L}_{S}(\mu, \sigma)$ with $k$-point distributions. The problem  $ \sup_{F \in \mathcal{L}_{k, S}(\mu, \sigma)}\E^{F}[(X-t)_{+}^2] $ is a  finite-dimensional  optimization problem and it can be  solved through  detailed mathematical analysis. The detailed  proof of Theorem \ref{th-2nd-TVS-S}   is given 
 in the appendix. 
 
 Theorem   \ref{th-2nd-TVS-S} provides  the explicit expression for the worst-case target semi-variance over the uncertainty set 
 $ \mathcal{L}_{S}(\mu, \sigma) $ with symmetric distributions. In Section \ref{sec:ER-TSV}  of this paper,  
 we propose a portfolio selection model based on  mean-target semi-variance with symmetric distributions   (M-TSV-S). This model  incorporates the symmetric features of daily stock  losses  and  aims to  minimize 
  the  the worst-case target semi-variance of portfolio loss  over undesirable scenarios.  As illustrated in 
  Section \ref{sec:ER-TSV}, this M-TSV-S model   can be reformulated as 
    a    minimization  problem, which 
 minimizes the   worst-case target semi-variance of  portfolio loss  over the uncertainty set 
 $ \mathcal{L}_{S}(\mu, \sigma)$.

\section{Worst-case target semi-variances   over undesirable scenarios}
\label{sec:WC-ER}

 In this section, for given  $t \in \R$ and $\lambda >0$, we  study the following optimization problem:
\begin{equation}
\label{pro:budget-l}  
\begin{aligned}
\sup_{F \in \mathcal{L}_{\lambda}}~&~ \E^{F}[(X-t)^2_+],
  \end{aligned}
 \end{equation}
where  ${\cal L}_{\lambda}$ is one of the uncertainty sets  $\mathcal{L}_{\lambda}(\mu, \sigma)$, $\mathcal{L}_{S,\lambda}(\mu, \sigma)$, and $\mathcal{L}^{+}_{\lambda}(\mu, \sigma)$ defined in \eqref{eq:budget<lambda}-\eqref{L+<lambda}.  Note that by Jensen's inequality,   
the condition $\E^{F}[(X-t)_-] \le \lambda$ in the set ${\cal L}_{\lambda}$ means that it must hold 
$\lambda \ge (\mu-t)_-$. Otherwise, if $\lambda < (\mu-t)_-$,
 then the set ${\cal L}_{\lambda}$ is empty and  
$\sup_{F \in \mathcal{L}_{\lambda}} \E^{F}[(X-t)^2_+] = \sup  \emptyset=-\infty$.  To exclude this trivial case, in problem  \eqref{pro:budget-l}, we assume 
$\lambda \ge (\mu-t)_- $ 
in the following discussion.

\subsection{Arbitrary or nonnegative  distributions} 

In this subsection, we first solve problem \eqref{pro:budget-l} with 
  ${\cal L}_\lambda = \mathcal{L}_{\lambda}(\mu, \sigma)$ and present the solutions to the problem in the following theorem.  The  proof  is provided   in the appendix.

    
      \begin{theorem}
\label{th:constraint}
  For $ \mu, t \in \R$, $\sigma>0$, 
 $\lambda >0$, and $ \lambda \ge (\mu-t)_-$,       we have
  \begin{align}
  \sup_{F \in \mathcal{L}_{\lambda}(\mu, \sigma)} \, \E^{F}[(X-t)^2_+]
              & =   \begin{cases}
             \sigma^2 + (\mu-t)_+^2, &  \lambda > (\mu-t)_-,  \\
            0, &  \lambda=(\mu-t)_-.
   \end{cases} 
       \label{eq:constraint}  
   \end{align}
 \end{theorem}
   
 Theorem \ref{th:constraint} provides the explicit expression for the worst-case target 
  semi-varaince over the uncertainty set $\mathcal{L}_{\lambda}(\mu, \sigma)$. This   expression 
  will be used to solve the  portfolio selection problem proposed in Section  \ref{sec:ER-TSV},  
   which is based on the expected excess profit-target semi-variance (EEP-TSV). In addition,  
    based on Theorem \ref{th:constraint} and its proof, we can 
 solve problem \eqref{pro:budget-l} with 
  ${\cal L}_\lambda = \mathcal{L}^+_{\lambda}(\mu, \sigma)$.
 First we give  an equivalent condition for the non-empty of the uncertainty set $ \mathcal{L}^+_{\lambda}(\mu, \sigma)$.
 
 \begin{proposition} \label{Prop:1}
Let  $\mu>0$,  $\sigma>0$, $\lambda>0$,  and $ \lambda \ge (\mu-t)_-$. 
 Then, the set $\mathcal{L}^+_{\lambda}(\mu, \sigma)$ is non-empty if and only if  $\lambda > (\mu - t)_-$ or $\lambda = (\mu - t)_-$ and $\sigma^2 \le \mu (t - \mu)$. 
 \end{proposition}

 In the following corollary we only consider  the case that the set $\mathcal{L}^+_{\lambda}(\mu, \sigma)$ is not an empty set since otherwise $\sup_{F \in \mathcal{L}^+_{\lambda}(\mu, \sigma)} \E^{F}[(X-t)_{+}^2]=-\infty$. The proof of this corollary is given in the appendix. 
 \begin{corollary}
\label{coro:positive<m}   
Let  $ t\in \R$, $ \mu >0$, $\sigma >0$,  $\lambda >0$, and $ \lambda\ge (\mu-t)_-$. If  $\lambda > (\mu - t)_-$ or $\lambda = (\mu - t)_-$ and $\sigma^2 \le \mu (t - \mu)$, 
then
  \begin{align*}
 \sup_{F \in \mathcal{L}^+_{\lambda}(\mu, \sigma)} \E^{F}[(X-t)_{+}^2]
          &\  = \ \sup_{F \in \mathcal{L}_{\lambda}(\mu, \sigma)} \E^{F}[(X-t)_{+}^2]  \ =   \begin{cases}
             \sigma^2 + (\mu-t)_+^2, &  \lambda > (\mu-t)_-,  \\
            0, &  \lambda=(\mu-t)_-, ~\sigma^2  \le   \mu (t-\mu).
   \end{cases}    
\end{align*}
\end{corollary}

In addition, note that the sets  $\mathcal{L}(\mu, \sigma) $ and $\mathcal{L}^+(\mu, \sigma) $ are the limiting cases 
 of $\mathcal{L}_{\lambda}(\mu, \sigma) $ and $\mathcal{L}^+_{\lambda}(\mu, \sigma)$, respectively,  as $\lambda \to \infty$. In fact, $\mathcal{L}(\mu, \sigma) =  \mathcal{L}_{\infty}(\mu, \sigma)$ and $\mathcal{L}^+(\mu, \sigma) = 
 \mathcal{L}^+_{ \infty}(\mu, \sigma)$. Thus, by setting $\lambda=\infty$ in Theorem \ref{th:constraint} and  Corollary \ref{coro:positive<m}, we immediately obtain the following corollary.
 \begin{corollary}
\label{coro:positive}
For $t\in \R$, $ \mu >0$, and $\sigma > 0$, we have
\begin{align}
 \sup_{F \in \mathcal{L}^+(\mu, \sigma)}  \E^{F}[(X-t)^2_+]
         & \ =   \  \sup_{F \in \mathcal{L}(\mu, \sigma)} \E^{F}[(X-t)_{+}^2] \ = \  \sigma^2 + (\mu-t)_+^2.  \label{eq:constraint-g}    
  \end{align}
\end{corollary} 
 
 The above corollaries imply that the worst-case value  of $ \E^{F}[(X-t)_{+}^2]$ over the set 
 $\mathcal{L}(\mu, \sigma)$  (resp. $\mathcal{L}_\lambda(\mu, \sigma))$  is the same as the worst-case value  of $ \E^{F}[(X-t)_{+}^2]$ over the set 
 $\mathcal{L}^+(\mu, \sigma) $ (resp. $\mathcal{L}^+_\lambda(\mu, \sigma))$. 
 
\subsection{Symmetric distributions} 

In this subsection, we solve problem \eqref{pro:budget-l} with  
  ${\cal L}_\lambda = \mathcal{L}_{\lambda,S}(\mu, \sigma)$,  which  is the problem 
  \begin{align} \label{pro:budget<m-sym-m1} 
  \sup_{F \in \mathcal{L}_{S,\lambda}(\mu, \sigma)} & ~~ \E^{F}[(X-t)^2_+].
  \end{align}
   For this problem,  the methods and ideas used in \cite{J77} and \cite{CHZ11}   do not work for problem \eqref{pro:budget<m-sym-m1}.    In this section,  we first  show 
problem      \eqref{pro:budget<m-sym-m1}   can be  reformulated  as a finite-dimensional optimization problem and then 
derive the explicit  and closed-form expression for the worst-case target semi-variance by solving the finite-dimensional optimization problem.

 \begin{proposition} \label{Prop:2}
Let  $\mu>0$,  $\sigma>0$, $\lambda>0$,  and $ \lambda \ge (\mu-t)_-$.  
Then, the set $\mathcal{L}_{S, \lambda}(\mu, \sigma)$ is non-empty if and only  
$\lambda > (\mu - t)_-$ or  $\lambda = (\mu - t)_-$ and $\sigma  >    t-\mu$. 
\end{proposition}

Note that if the set $\mathcal{L}_{S,\lambda}(\mu, \sigma)$ is non-empty under the condition of $\lambda = (\mu - t)_-$ and $\sigma  >    t-\mu$, 
 then we have $\sup_{F \in \mathcal{L}_{S,\lambda}(\mu, \sigma)} \mathbb{E}^{F}[(X - t)^2_+] = 0$. In the following results, we will only consider the non-trivial case where $\lambda > (\mu - t)_-$. 

 \begin{lemma}\label{eq:0919-1}
 For any $t \in \R$,  $ \mu >0$,    $\sigma >0$,  
  and $ \lambda > (\mu-t)_-$, problem  \eqref{pro:budget<m-sym-m1} 
 is equivalent to the problem
 \begin{equation}
\label{pro:budgetapp<m-sym-k}
\begin{aligned}
\sup_{F \in \mathcal{L}_{6,S,\lambda}(\mu, \sigma)} & ~~ \E^{F}[(X-t)^2_+],
  \end{aligned}
  \end{equation}
  where
$
 \mathcal{L}_{6, S,\lambda}(\mu, \sigma) = \big \{F \in \mathcal{L}_{S,\lambda}(\mu, \sigma): F \ \mbox{is a six-point distribution}\}.
$
     \end{lemma}

Problem \eqref{pro:budgetapp<m-sym-k}  is  a finite-dimensional problem and is proved  in the appendix.
  By solving this finite-dimensional problem, we obtain the explicit solution to 
 problem  \eqref{pro:budget<m-sym-m1}, which is presented in the following theorem and also   proved in the appendix. 
 
 \begin{theorem} 
 \label{th-2nd-TVS-S-m}
 For $ t\in \R$, $ \mu >0$,    $\sigma >0$,   and   $ \lambda > (\mu-t)_-$, define  $m=\lambda -\mu+t$. Then,  $m>0$ and following results hold:

\begin{itemize}
    \item[(a)] If $\sigma \le m$, we have    
  \begin{equation*}
\sup_{F \in \mathcal{L}_{S, \lambda}(\mu, \sigma)}\E^{F}[(X-t)_{+}^2]=
        \begin{cases}
    \sigma^2 + (t-\mu)^2, & \mu -m\le t \le \mu - \sigma,\\
    \frac{1}{2}(\mu- t+\sigma)^2, & \mu - \sigma < t \le \mu,\\
    \frac{1}{2} \sigma^2, & t > \mu.
    \end{cases}
\end{equation*}
\item[(b)] If $m <\sigma \le 2m$, we have
\begin{equation*}
\sup_{F \in \mathcal{L}_{S, \lambda}(\mu, \sigma)}\E^{F}[(X-t)_{+}^2]=
        \begin{cases}
    \frac{1}{2}\sigma^2 + 2m(\mu - t) - \frac{(t-\mu)^2}{2}, & \mu - m \le t \le \mu + \sigma -2m,\\
    \frac{1}{2}(\mu- t+\sigma)^2, & \mu + \sigma -2m < t \le \mu,\\
    \frac{1}{2} \sigma^2, & t > \mu.
    \end{cases}
\end{equation*}
  
    \item[(c)] If $\sigma > 2m$, we have    
  \begin{equation*}
\sup_{F \in \mathcal{L}_{S, \lambda}(\mu, \sigma)}\E^{F}[(X-t)_{+}^2]=
        \begin{cases}
    \frac{1}{2}(\mu- t+\sigma)^2, & \mu -m\le t \le \mu,\\
    \frac{1}{2} \sigma^2, & t > \mu.
    \end{cases}
\end{equation*}
  \end{itemize} 
 
  \end{theorem}

 Theorem \ref{th-2nd-TVS-S-m} provides  the explicit expression for the worst-case target semi-variance over the uncertainty set 
 $ \mathcal{L}_{S, \lambda}(\mu, \sigma) $ with symmetric distributions. In Section \ref{sec:ER-TSV}  of this paper,  
 we propose a portfolio selection model based on 
 expected excess profit-target semi-variance with symmetric distributions  
   (EEP-TSV-S). This model   incorporates the symmetric features of daily stock  losses and   minimizes the  the worst-case target semi-variance of portfolio loss over undesirable scenarios.  It 
    can be transformed into to a minimization problem, which 
 minimizes the  the worst-case target semi-variance of portfolio loss  over the uncertainty set 
 $ \mathcal{L}_{S, \lambda}(\mu, \sigma)$. 
   
\section{Applications to robust portfolio selection}
\label{sec:ER-TSV}

In this section, we consider the applications of the worst-case  target semi-variances  derived in Sections \ref{WC-SD} and \ref{sec:WC-ER}  to robust  portfolio selection problems. 

Let  ${\bm X}^\top=(X_1,...,X_d) \in \R^d$ be a random vector representing the loss vector in an investment portfolio with $X_i$ being the loss in investing in an asset  $i$,  $ i=1,...,d$. The loss of the portfolio is  $\bm{w}^\top \bm{X} =\sum_{i=1}^d w_iX_i,$ where $\bm{w}^\top=(w_1,...,w_d) \in \R^d$ with $w_i$ being the allocation/selection on asset $i$. Without loss of generality, we assume that the investor's total initial wealth to be invested is 1, 
which means  that $\bm{w}^\top\bm{e} = 1$,  where $\bm{e}^\top=(1,...,1)$ is a $d$-dimensional unit vector.
 In this section, 
 we use the set of portfolios  $\mathcal{W} = \{ \bm{w} \in \R^d: \bm{w}^\top\bm{e} = 1  \}$ to 
 denote  that an investor may short sell a stock, and use the set of portfolios 
 $\mathcal{W}^+ = \{ \bm{w} \in \R^d: \bm{w}^\top\bm{e} = 1, \, \bm{w} \geq  0 \}$  to 
 indicate   that an investor does not  short sell any stock or to denote the rule that   short-selling
 is not allowed.

               In the classical mean-variance (M-V) model, 
      the mean vector $\bm{\mu}$ and covariance matrix ${\bm \Sigma}$ of the loss vector ${\bm X}$   are assumed to 
       be known, which essentially assume that  the `true' (joint) distribution $G$ of  
       $\bm X$ is unknown   and  only the  
        mean vector $\bm{\mu}$ and covariance matrix ${\bm \Sigma}$ of  ${\bm X}$ are available. 
        In other words, the possible   (joint) distribution $G$ of  
       $\bm X$ is assumed to belong to the following set of distributions:          
    \begin{equation}
    \label{M}
   \mathcal{M}(\bm{\mu}, \bm{\Sigma}) = \big \{ G \in  {\mathcal F}(\R^d): \, \E[\bm{X}] = \bm{\mu}, \, \cov[\bm{X}] 
   = {\bf \Sigma} \big \}, 
  \end{equation}   
 where  ${\mathcal F}(\R^d)$ is the set of all $d$-dimensional distributions defined on $\R^d$. 
For any  $G \in  \mathcal{M}(\bm{\mu}, \bm{\Sigma})$,     $\E^G[{\bm w}^\top {\bm X} ] =  {\bm w}^\top \bm{\mu}$ and  ${\rm Var}^G({\bm w}^\top {\bm X}) = {\bm w}^\top {\bm \Sigma}  \bm{w}$.
   The classical   M-V  portfolio selection model  can be  formulated as
\begin{equation}
\label{C-MV-model}
\begin{aligned}
 &\min_{\bm{w} \in \mathcal{W}}  \, \sup_{G \in  \mathcal{M}(\bm{\mu}, \bm{\Sigma})    }\, \mbox{Var}^G(\bm{w}^{\top}\bm{X}) =   \min_{\bm{w} \in  \mathcal{W}}  ~  \bm{w^{\top}}{\bf \Sigma}\bm{w} \\
     & {\rm s.t.}  ~~ 
        \bm{w^{\top}\mu} \le \nu,
\end{aligned}
\end{equation} 
where $
\bm{w^{\top}\mu} \le \nu$ is equivalent to $
    -\bm{w^{\top}\mu} \ge -\nu$ that 
represents  a constraint on the expected return of the portfolio.   However,  if the distribution $G$ of ${\bm X}$ is uncertain and belongs to 
  $\mathcal{M}(\bm{\mu}, \bm{\Sigma})$,     then  the expected downside loss or expected regret  $\E^G[({\bm w}^\top {\bm X} -t)_+]$  and the target semi-variance $\E^G[({\bm{w}}^\top {\bm X} -t)^2_+]$ are also uncertain. 
   To incorporate the symmetric information of loss vectors and    minimize the worst-case  
  target semi-variance of the portfolio loss,  we propose the following two target semi-variance (TSV)-based robust   portfolio 
selection models:   
\begin{itemize}
\item[(1)] Portfolio 
selection model based on  the mean-target semi-variance with symmetric distributions (M-TSV-S), which   is formulated as 
\begin{equation}
\label{pro:pf-m-EL-g}
\begin{aligned}
    &  \min_{\bm{w} \in \mathcal{W}} \ \sup_{G \in   \mathcal{M}_S(\mu, \sigma)}\, \E^{G}[(\bm{w}^\top\bm{X} - t)^2_+]\\
    & {\rm s.t.} \quad  
   \bm{w^{\top}\mu} \le \nu, 
    \end{aligned}
\end{equation}
where
 \begin{equation}
 \label{MS} 
\mathcal{M}_S(\bm{\mu}, \bm{\Sigma}) = \big \{ G \in  \mathcal{M}(\bm{\mu}, \bm{\Sigma}):  \,  \mbox{$G$ is symmetric} \big \}
\end{equation} 
and  $\nu$ is a risk level or equivalently $-\nu$ is a desirable   minimum level for the expected return.

\item[(2)] Portfolio 
selection model  
 minimizing   the worst-case   target semi-variance of portfolio loss 
  over undesirable scenarios, which is    formulated as 
\begin{equation}
\label{pro:pf-m-EL<lambda}
\begin{aligned}
    &  \min_{\bm{w} \in \mathcal{W}^+} \ \sup_{G \in   \mathcal{M}_{\bm{w},  \lambda}}\, \E^{G}[(\bm{w}^\top\bm{X} - t)^2_+]\\
    \end{aligned}
\end{equation}
where 
\begin{equation}
\label{Mwl}
 \mathcal{M}_{\bm{w},   \lambda}  =\big \{ G \in  \mathcal{M}:  \,  
  \E^G[(\bm{w}^{\top}\bm{X} -t)_-]  \le \lambda \big \}, 
 \end{equation} 
the uncertainty set  $\mathcal{M}$ is one of  $\mathcal{M}(\bm{\mu}, \bm{\Sigma})$ and  $\mathcal{M}_S(\bm{\mu}, \bm{\Sigma})$, and $\lambda >0$ is a  desirable minimum  level 
for the expected excess profit over the target return $-t$.
\end{itemize} 
We solve problems \eqref{pro:pf-m-EL-g} and  \eqref{pro:pf-m-EL<lambda} in the following two subsections.

\subsection{Robust  portfolio selection with  symmetric  distributions}  

In this subsection, we solve problem \eqref{pro:pf-m-EL-g}.  We first  give the  definition of symmetric random vector  or multivariate symmetric distribution.  To do so,  denote the set of all Borel measurable sets in $\R^d$ by  $ \mathfrak{B}(\R^d)$. For a set $A \subset \R^d$ and a  vector $\bm{a} \in \R^d$,  denote $-A$ by $-A=\{\bm{x} \in \R^d: \, -\bm{x} \in A\}$ and denote $A- \bm{a}$ by  $A- \bm{a} =\{\bm{x}-  \bm{a} \in \R^d: \, \bm{x} \in A\}$.

\begin{definition}
\label{sdf-m}
A random vector  $\bm{X} \in \R^d$ or  its  joint  distribution $G$   is said to be symmetric if there exists a vector  $\bm{a} \in \R^d$ such that $ \mathbb{P}
(\bm{X} - \bm{a} \in B ) = \mathbb{P}(\bm{X}-\bm{a}  \in -B)$ under the distribution $G$, for all $B \in \mathfrak{B}(\R^d)$. If such a vector $\bm{a}$ exists, random vector  $\bm{X}$ or its distribution is said to be symmetric about   $\bm{a}$. \hfill $\Box$
   \end{definition}

Intuitively, random vector $\bm{X}$  is symmetric about  $\bm{a}$ if and only if  $\bm{X}-\bm{a}$ is  symmetric about   the origin of $\R^d$. Examples of  continuous multivariate symmetric distributions include multivariate normal distributions, multivariate  $t$-distributions,  multivariate  elliptical distributions, and many others. In addition, a constant random vector  is also symmetric according to Definition \ref{sdf-m}. The proof of the following result is straightforward and thus omitted.

 \begin{lemma}
\label{le-linear-S}
(i) If random vector $\bm{X} \in \R^d$ or its distribution $G$ is symmetric about  $\bm{a}$, then for any  vector $\bm{w} \in \R^d$,  the distribution of $\bm{w}^\top \bm{X}$ is symmetric about   $\bm{w}^\top \bm{a}$.

(ii) If $d$-dimensional random vectors   $\bm{X}_1$ and $\bm{X}_2$ are independent and symmetric about  $\bm{\mu}_1$ and $\bm{\mu}_2$, respectively,
then $\bm{X}_1+\bm{X}_2$ is symmetric about  $\bm{\mu}_1+\bm{\mu}_2$.
\end{lemma}


We now  denote  the the multivariate mean-covariance uncertainty set with symmetric distributions  by  $\mathcal{M}_S(\bm{\mu}, {\bf \Sigma})$ that is defined 
in \eqref{MS}.
  Moreover, for a given $\bm{w} \in \R^d$, define  $\mathcal{L}_{\bm{w}, S}(\bm{\mu}, {\bf \Sigma})$ as the one-dimensional distribution set generated from  the  distribution of 
$\bm{w}^\top \bm{X}$ when the joint distribution of 
$\bm{X}$ belongs to  $\mathcal{M}_S(\bm{\mu}, {\bf \Sigma})$, namely
 \begin{align}
  \mathcal{L}_{\bm{w}, S}(\bm{\mu}, {\bf \Sigma}) &= \left\{ F_{\bm{w}^\top\bm{X}} \in \mathcal{F}(\R):
    \,\text{The joint distribution $G$ of} \ \bm{X} \  \text{belongs to} \  \mathcal{M}_S(\bm{\mu}, {\bf \Sigma}) \right\}.  \label{eq:mwS}
         \end{align}
        \begin{lemma}
\label{lem:multi-S}  
If  the   covariance matrix ${\bf \Sigma}$ of the loss  random vector  $\bm{X}$ is positive definite  and $\bm{w} \neq \bm{0}$, then 
\begin{equation}
\mathcal{L}_{{\bm{w}},S}(\bm{\mu}, {\bf \Sigma}) = \mathcal{L}_S\big (\bm{w^{\top}\mu}, \sqrt{\bm{w^{\top}}{\bf \Sigma}\bm{w}} \, \big ) \label{Lw,m=Lm-S}
\end{equation} 
and
 \begin{equation}
 \label{sup-Lw,m=Lm-S}
 \sup_{G \in   \mathcal{M}_{S}(\bm{\mu}, {\bf \Sigma})   }\, \E^{G}[(\bm{w}^{\top}\bm{X} - t)^2_+] 
=\sup_{F \in \mathcal{L}_S\left(\bm{w^{\top}\mu}, \, \sqrt{\bm{w^{\top}}{\bf \Sigma}\bm{w}} \, \right)     }\, \E^{F}[(X - t)^2_+], 
 \end{equation}
where  set
 $\mathcal{L}_S(\mu, \sigma)$ is   defined in  \eqref{LS} for any $\mu \in \R$ and any $\sigma \in \R^+$,  and $X$ is a random variable with  a distribution belonging to   $\mathcal{L}_S\big (\bm{w^{\top}\mu}, \sqrt{\bm{w^{\top}}{\bf \Sigma}\bm{w}}\, \big )$.
\end{lemma}
\begin{proof} For any distribution $F_{\bm{w}^\top\bm{X}} \in  \mathcal{L}_{\bm{w}, S}(\bm{\mu}, {\bf \Sigma}) $ with the joint distribution $G$  of $ \bm{X} $  belonging  to $  
      \mathcal{M}_S(\bm{\mu}, {\bf \Sigma})$, we have $ \E^G[ \bm{w}^\top \bm{X}] =\bm{w}^\top \bm{\mu}$ and 
       $ \cov^G[\bm{X}]= \bm{w}^\top {\bf \Sigma} \bm{w}$. In addition, by Lemma \ref{le-linear-S}(i), we see that  $\bm{w}^\top \bm{X}$ is symmetric as $ \bm{X}$ is symmetric. Hence,  $ F_{\bm{w}^\top\bm{X}} \in \mathcal{L}_S\big (\bm{w^{\top}\mu}, \sqrt{\bm{w^{\top}}{\bf \Sigma}\bm{w}} \, \big )$.  Thus,   $\mathcal{L}_{\bm{w},S}(\bm{\mu}, {\bf \Sigma}) \subseteq \mathcal{L}_S\big (\bm{w^{\top}\mu}, \sqrt{\bm{w^{\top}}{\bf \Sigma}\bm{w}}\, \big )$. Next, 
      we prove $ \mathcal{L}_S\big (\bm{w^{\top}\mu}, \sqrt{\bm{w^{\top}}{\bf \Sigma}\bm{w}}\, \big ) \subseteq \mathcal{L}_{\bm{w}, S}(\bm{\mu}, {\bf \Sigma})$.      
 Similar to the proof of \citet[Lemma 2.4]{CHZ11}, for      $\bm{w}\not=0 \in \R^d$  and 
any  distribution  $F  \in \mathcal{L}_S\big (\bm{w^{\top}\mu}, \sqrt{\bm{w^{\top}}{\bf \Sigma}\bm{w}}\, \big )$, we 
construct  a $d$-dimensional  random vector  $\bm{X}^*$ as 
\begin{equation*}
\bm{X} ^* =  \frac{((\bm{w}^{\top}{\bf \Sigma} \bm{w}){\bf \Sigma} - {\bf \Sigma} w w^\top {\bf \Sigma})^{\frac{1}{2}} \bm{Z} }{\sqrt{\bm{w}^{\top}{\bf \Sigma} \bm{w}}} + \frac{(Y - \bm{w}^{\top} \bm{\mu}) {\bf \Sigma}
\bm{w}}{\bm{w}^{\top}{\bf \Sigma} \bm{w}} + \bm{\mu}, 
\end{equation*}
where $Y$ is a random variable with the distribution $F$ and  
 $\bm{Z}$ is a $d$-dimensional standard normal random vector  independent  of  $Y$. Then, 
  $\bm{w}^{\top} \bm{X}^* = Y$,  $\E[\bm{X}^*] = \bm{\mu}$, and $\cov[\bm{X}^*] = {\bf \Sigma}$.  
 In addition, by Lemma \ref{le-linear-S}(i) and (ii), we see that $\bm{X}^*$ is symmetric and thus 
  $\bm{w}^{\top} \bm{X}^* = Y$ is symmetric as well.   Hence, 
 the joint distribution  of $ \bm{X}^*$  belongs to $    \mathcal{M}_S(\bm{\mu}, {\bf \Sigma})$ and the distribution $F$ of  
 $\bm{w}^{\top} \bm{X}^* = Y$ belongs to $\mathcal{L}_{{\bm{w}},S}(\bm{\mu}, {\bf \Sigma})$, which 
 mean that     $ \mathcal{L}_S\big (\bm{w^{\top}\mu}, \sqrt{\bm{w^{\top}}{\bf \Sigma}\bm{w}} \, \big ) \subseteq \mathcal{L}_{\bm{w}, S}(\bm{\mu}, {\bf \Sigma})$. Therefore,  we conclude that $\mathcal{L}_{\bm{w}, S}(\bm{\mu}, {\bf \Sigma}) = \mathcal{L}_S\big (\bm{w^{\top}\mu}, \sqrt{\bm{w^{\top}}{\bf \Sigma}\bm{w}} \, \big )$. 
 It is obvious that 
  $
 \sup_{G \in   \mathcal{M}_{S}(\bm{\mu}, {\bf \Sigma})   }\, \E^{G}[(\bm{w}^{\top}\bm{X} - t)^2_+] 
=\sup_{F \in \mathcal{L}_{\bm{w}, S}(\bm{\mu}, {\bf \Sigma})   }\, \E^{F}[(X - t)^2_+],
 $
 which,  together with   \eqref{Lw,m=Lm-S}, implies \eqref{sup-Lw,m=Lm-S}.  
 \end{proof}

To better present  the optimal portfolio selections   derived in this paper, we define parameters    $u$, $v_0$, $v_1$, $v_2$ as follows:
\begin{align}
\label{v}
u &= (\bm{e}^{\top} {\bf \Sigma}^{-1}\bm{e})(\bm{\mu}^{\top} {\bf \Sigma}^{-1}\bm{\mu}) - (\bm{e}^{\top} {\bf \Sigma}^{-1}\bm{\mu})^2, \ \,
    v_0 = \frac{\bm{e}^{\top} {\bf \Sigma}^{-1}\bm{e}}{u}, \  \, v_1 = \frac{\bm{e}^{\top} {\bf \Sigma}^{-1}\bm{\mu}}{u}, \  \, v_2 = \frac{\bm{\mu}^{\top} {\bf \Sigma}^{-1}\bm{\mu}}{u}, 
\end{align}
where $\bm{w}, \bm{\mu} \in \R^d$ with $\bm{w}^{\top}\bm{e} = 1$, and  ${\bf \Sigma}$ is a  $d \times d$  positive definite matrix. Note that for any $\bm{\mu} \in \R^d$, it holds $u \ge 0$ since ${\bf \Sigma}$ is  a positive definite matrix.  
However, to guarantee that the optimal solutions exist, 
   we assume that $\bm{\mu}$ and $\bm{e}$ are not linearly dependent, or equivalently, 
  assume that     for any $c \in \R$, $\bm{\mu} \not= c \, \bm{e}$. This assumption 
  implies  $u>0$ and is also used in the classical M-V  portfolio  selection model.

Moreover,  to present the optimal solution to problem \eqref{pro:pf-m-EL-g},
 we define 
 $$h_{S,t}(\mu, \sigma)
= \sup_{F \in \mathcal{L}_S(\mu, \sigma)}\E^F[(X-t)^2_+].$$ 
 By 
  Theorem \ref{th-2nd-TVS-S}, we  can write  $h_{S,t}(\mu, \sigma)$  as a function of $\sigma$ 
with  the following expression: 
    \begin{itemize}
\item[(i)] If $\mu > t$, then 
    \begin{equation}
   \label{upm2S-sigma-1}
h_{S,t}(\mu, \sigma) =  
        \begin{cases}
    \sigma^2 + (t-\mu)^2, & 0 < \sigma  \le  \mu-t,\\
    \frac{1}{2}(\mu- t+\sigma)^2, & \sigma > \mu-t.
    \end{cases}
\end{equation}

\item[(ii)] If $\mu \le t$, then  
 \begin{equation}
   \label{upm2S-sigma-2}
   h_{S,t}(\mu, \sigma)  = 
  \frac{\sigma^2}{2},    \ \ \sigma>0. 
\end{equation} 
  \end{itemize} 
In addition, define 
\begin{align}
\xi_1^* & = \argmin_{\xi \le t} \ h_{S, t}\big (\xi, \, \sqrt{v_0 \xi^2-2 v_1 \xi +v_2} \, \big ), \label{xi1*} \\
  h^1_{S, t}(\xi_1^*) &= h_{S, t}\big (\xi_1^*, \, \sqrt{v_0 (\xi_1^*)^2-2 v_1 \xi _1^*+v_2} \, 
  \big )=\frac{1}{2} 
\big (v_0 (\xi_1^*)^2-2 v_1 \xi_1^* +v_2 \big ), \label{h1} \\
\xi_2^* & = \argmin_{t \le \xi \le v} \ h_{S, t}\big (\xi, \, \sqrt{v_0 \xi^2-2 v_1 \xi +v_2}\, \big ),  \label{xi2*}\\
h^2_{S, t}(\xi_2^*) & = h_{S, t}\big (\xi_2^*, \, \sqrt{v_0 (\xi_2^*)^2-2 v_1 \xi_2^* +v_2} \, \big ). \label{h2} 
  \end{align} 
Note that   by  \eqref{upm2S-sigma-2},   
 \begin{equation} 
 \label{xi1*-sol}
 \xi_1^*=\argmin_{\xi \le t}  \frac{1}{2} \big ( v_0 \xi^2-2 v_1 \xi +v_2 \big ) 
 = \min\big \{ \frac{v_1}{v_0}, \, t \big \}. 
 \end{equation} 
By  \eqref{upm2S-sigma-1}, if $t < \nu$, 
we see that $h_{S, t}\big (\xi, \, \sqrt{v_0 \xi^2-2 v_1 \xi +v_2} \, \big )  $ is a continuous function of $\xi$ on $[t, \nu]$. 
Thus, there exists $\xi_2^* \in [t, \nu]$ such  that
$
 \min_{ t \le  \xi \le  \nu} 
 \  h_{S, t}\big (\xi, \, \sqrt{v_0 \xi^2-2 v_1 \xi +v_2} \, \big ) 
  = h^2_{S, t}(\xi_2^*).$

\begin{proposition}
\label{prop:upm-S}
Assume the covariance matrix  ${\bf \Sigma}$ of the loss random vector $\bm{X}$ is  positive definite. 
Then,  problems \eqref{pro:pf-m-EL-g} 
 has a unique solution $\bm{w}_{S, \nu,t}^*$  that
 has the following expression:
\begin{equation}
\label{w*2}
    \bm{w}_{S, \nu,t}^* = ({\bf \Sigma}^{-1}\bm{\mu},  \quad {\bf \Sigma}^{-1}\bm{e})
    \begin{pmatrix}
    v_0 & -v_1\\
    -v_1 & v_2
    \end{pmatrix}
    \begin{pmatrix}
\xi^*
\\
    1
    \end{pmatrix}.
\end{equation}
Here, $\xi^*$ in \eqref{w*2} has the following expressions: 
  \begin{itemize}  
\item[(i)]      If $t \ge \nu$, then $\xi^*
= \min\big \{ \frac{v_1}{v_0}, \, \nu \big \}$.

\item [(ii)]  If $t < \nu$ and 
$h^1_{S, t}(\xi_1^*)  \le h^2_{S, t}(\xi_2^*)$,
 then $\xi^*
= \min\big \{ \frac{v_1}{v_0}, \, t \big \}$.

\item[(iii)] 
If $t < \nu$ and 
$h^1_{S, t}(\xi_1^*)  > 
 h^2_{S, t}(\xi_2^*)$,
  then $\xi^* =\xi_2^*
=\argmin_{ t \le  \xi \le  \nu} 
 \,  h_{S, t}\big (\xi, \, \sqrt{v_0 \xi^2-2 v_1 \xi +v_2}\, \big )$. 

\end{itemize} 

\end{proposition}
\begin{proof} By Lemma \ref{lem:multi-S}, for the positive definite  matrix ${\bf \Sigma}$ and $\bm{w} \neq \bm{0}$, the inner optimization  problem of \eqref{pro:pf-m-EL-g}     reduces to 
the 
problem 
$\sup_{F \in \mathcal{L}_{S}\big (\bm{w^{\top}\mu}, \, \sqrt{\bm{w^{\top}}{\bf \Sigma}\bm{w}}\, \big ) }\, \E^{F}[(X - t)^2_+]= h_{S, t}\big (\bm{w^{\top}\mu}, \sqrt{\bm{w^{\top}}{\bf \Sigma}\bm{w}}\, \big ),$
which  has been solved in   Theorem \ref{th-2nd-TVS-S}.     Hence, 
  problem \eqref{pro:pf-m-EL-g}    is equivalent to the following  problem: 
   \begin{equation}
\label{pro:pf-S-2}
\begin{aligned}
    \min_{\bm{w} \in \mathcal{W}} & ~~  h_{S,t}\big (\bm{w^{\top}\mu}, \sqrt{\bm{w^{\top}}{\bf \Sigma}\bm{w}} \, \big )~~
     {\rm s.t.} & ~~  
     \bm{w^{\top}\mu} \le \nu, 
\end{aligned}
\end{equation}
this is again equivalent to
   \begin{equation*}
\begin{aligned}
    \min_{\xi\in\R,\,\bm{w} \in \mathcal{W}} & ~~  h_{S,t}\big (\xi, \sqrt{\bm{w^{\top}}{\bf \Sigma}\bm{w}}\, \big ), ~~
     {\rm s.t.}  & ~~   
     \bm{w^{\top}\mu} =\xi\le \nu, 
\end{aligned}
\end{equation*}
which can be expressed as the following problem: 
   \begin{equation}
       \begin{aligned}
\label{eq:mve-S}
&
\min_{\xi \in \R} \ \min_{\bm{w} \in \R^d, \, \bm{w}^{\top} \bm{e} = 1, \,  \bm{w^{\top}\mu} = \xi} \ \, h_{S,t}\big (\xi, \sqrt{\bm{w^{\top}}{\bf \Sigma}\bm{w}}\, \big ), 
  & {\rm s.t.} \quad    \xi \le  \nu. 
 \end{aligned}    
     \end{equation}
 By \eqref{upm2S-sigma-1} and \eqref{upm2S-sigma-2},    it is easy to see  that for any  $\xi \in \R$, $h_{S,t}(\xi, \sqrt{\sigma^2})$  is increasing in $\sigma^2$. 
 Therefore,
 \begin{equation}
 \min_{\bm{w} \in \R^d, \,   \bm{w}^{\top} \bm{e} = 1,  \, \bm{w^{\top}\mu} = \xi} h_{S,t}\big (\xi, \, \sqrt{\bm{w^{\top}}{\bf \Sigma}\bm{w}}\, \big ) = h_{S, t}\Big (\xi, \sqrt{\min_{\bm{w} \in \R^d, \,   \bm{w}^{\top} \bm{e} = 1,  \, \bm{w^{\top}\mu} = \xi}\bm{w^{\top}}{\bf \Sigma}\bm{w}} \, \Big ),
 \end{equation}    
 which means that 
 $$
 \bm{w}_{\xi}^* = \argmin_{\bm{w} \in \R^d, \,   \bm{w}^{\top} \bm{e} = 1,  \, \bm{w^{\top}\mu} = \xi} ~h_{S, t}\big (\xi, \sqrt{\bm{w^{\top}}{\bf \Sigma}\bm{w}} \, \big ) =  \argmin_{\bm{w} \in \R^d, \,   \bm{w}^{\top} \bm{e} = 1,  \, \bm{w^{\top}\mu} = \xi} ~\bm{w^{\top}}{\bf \Sigma}\bm{w}.
 $$
 It is well-known  that  
 \begin{equation*}
    \bm{w}^*_{\xi} \ = \ \argmin_{\bm{w} \in \R^d, \,   \bm{w}^{\top} \bm{e} = 1,  \, \bm{w^{\top}\mu} = \xi} ~\bm{w^{\top}}{\bf \Sigma}\bm{w}  \ = \  ({\bf \Sigma}^{-1}\bm{\mu} \quad {\bf \Sigma}^{-1}\bm{e})
    \begin{pmatrix}
    v_0 & -v_1\\
    -v_1 & v_2
    \end{pmatrix}
    \begin{pmatrix}
    \xi\\
    1
    \end{pmatrix}
\end{equation*} 
and $(\bm{w}^*_\xi)^{\top}{\bf \Sigma}\bm{w}^*_\xi=v_0 \xi^2-2 v_1 \xi +v_2$.
  Therefore, problem 
\eqref{eq:mve-S} is reduced to the following one-variance optimization  problem: 
   \begin{equation}
\label{pro:mv-case1-S}
\begin{aligned}
 \min_{\xi \le \nu} &\ \ h_{S, t}\big (\xi, \, \sqrt{v_0 \xi^2-2 v_1 \xi +v_2}\, \big ).
  \end{aligned}
 \end{equation}
 
   \begin{itemize}
\item[(a)]   If $t \ge   \nu$, by  \eqref{upm2S-sigma-2}, we have  
$
 \min_{\xi \le \nu} \ h_{S, t}\big (\xi, \, \sqrt{v_0 \xi^2-2 v_1 \xi +v_2}\, \big ) =   \min_{\xi \le \nu}  \, \frac{1}{2}\, 
 \big (v_0 \xi^2-2 v_1 \xi +v_2 \big ).$
Let $\xi^*=\argmin_{\xi \le \nu}  \frac{1}{2} \, \big ( v_0 \xi^2-2 v_1 \xi +v_2 \big )$. It is easy to see that $\xi^*
= \min\big  \{ \frac{v_1}{v_0}, \, \nu \big  \}$.   
\item[(b)]   If $t < \nu$, we have  
 \begin{equation*}
\begin{aligned}
 & ~ \min_{\xi < \nu} \ h_{S, t}\big (\xi, \, \sqrt{v_0 \xi^2-2 v_1 \xi +v_2}\, \big ) \\
 =& ~ \min\big \{\, \min_{\xi \le t} \  h_{S, t}\big (\xi, \, \sqrt{v_0 \xi^2-2 v_1 \xi +v_2}\, \big ), \  \ \min_{ t \le  \xi \le  \nu} 
 \  h_{S, t}\big (\xi, \, \sqrt{v_0 \xi^2-2 v_1 \xi +v_2}\, \big ) \big \}, 
  \end{aligned}
 \end{equation*}   
where
    by  \eqref{upm2S-sigma-2},   
 \begin{equation*}
\begin{aligned}
 \min_{\xi \le t} \ h_{S, t}\big (\xi, \, \sqrt{v_0 \xi^2-2 v_1 \xi +v_2}\, \big )& 
 = \frac{1}{2}\,  \min_{\xi \le t}  \, v_0 \xi^2-2 v_1 \xi +v_2 =\frac{1}{2} \left(v_0 (\xi_1^*)^2-2 v_1 \xi_1^* +v_2\right),
   \end{aligned}
 \end{equation*}
and
 $\xi_1^*=\argmin_{\xi \le t}  v_0 \xi^2-2 v_1 \xi +v_2 
= \min\big \{ \frac{v_1}{v_0}, \, t \big  \}$. In addition,  by  \eqref{upm2S-sigma-1}, 
we  see  that \\
$h_{S, t}\big (\xi, \, \sqrt{v_0 \xi^2-2 v_1 \xi +v_2} \, \big )  $ is a continuous function of $\xi$ on $[t, \nu]$. 
Thus, there exists $\xi_2^*$ such that
$ 
 \min_{ t \le  \xi \le  \nu} 
 \  h_{S, t}\big (\xi, \, \sqrt{v_0 \xi^2-2 v_1 \xi +v_2}\, \big ) =h_{S, t}\big (\xi_2^*, \, \sqrt{v_0 (\xi_2^*)^2-2 v_1 \xi_2^* +v_2}\, \big ).$
   \end{itemize}
By combining cases (a) and (b), we complete the proof.   
\end{proof}

\begin{remark} 
The optimal portfolio selection $\bm{w}^*_{\nu}$ to  the classical M-V    problem \eqref{C-MV-model}  
 has the following expression:
\begin{equation}
\label{w*-MV}
    \bm{w}_{\nu}^* = ({\bf \Sigma}^{-1}\bm{\mu},  \quad {\bf \Sigma}^{-1}\bm{e})
    \begin{pmatrix}
    v_0 & -v_1\\
    -v_1 & v_2
    \end{pmatrix}
    \begin{pmatrix}
   \min\big  \{\frac{v_1}{v_0},  \nu\big  \}\\
    1
    \end{pmatrix}.
\end{equation}
 Proposition \ref{prop:upm-S} shows that  if an investor  sets a  low  target return $-t$ or 
a  high threshold loss level $t$, say $t \ge \nu$, then    $\xi^*
= \min\big \{ \frac{v_1}{v_0}, \, \nu \big \}$ and     the optimal strategy  $ \bm{w}_{S, \nu,t}^* $   
is the same 
as the optimal strategy \eqref{w*-MV}
derived from  
  the classical M-V  model \eqref{C-MV-model}.   However, if an investor has a high return target $-t$ or a  low  threshold loss level $t$, say $t < \min\{\frac{v_1}{v_0}, \nu\}$, then 
   $\xi^*
= \min\big \{ \frac{v_1}{v_0}, \, t \big \} =t$ or  $\xi^* =\xi_1^*$
  and 
 the optimal strategy   $ \bm{w}_{S, \nu,t}^* $   
is different from   the optimal strategy \eqref{w*-MV}
derived from     the classical M-V  model \eqref{C-MV-model}.    As illustrated in the numerical experiments  given in Section \ref{sec:num} of this paper,  the portfolio performance with the   strategy derived from  \eqref{pro:pf-m-EL-g} 
  outperforms the   portfolio performance with 
   the strategy 
derived from   
  the classical M-V model \eqref{C-MV-model}.   
\hfill $\Box$

       \end{remark}

\subsection{Robust portfolio selection with constraint on expected regret}  

In this subsection, we solve problem \eqref{pro:pf-m-EL<lambda}.  First, we point  that by Jensen's inequality, for any  $\bm{w} \in \mathcal{W}^+$,  $ \E^{G}[(\bm{w}^\top\bm{X} - t)_-]  \ge (\bm{w}^\top\bm{\mu} - t)_-$
for all $G\in \mathcal M$ that  is $\mathcal{M}(\bm{\mu}, \bm{\Sigma})$ or   $\mathcal{M}_S(\bm{\mu}, \bm{\Sigma})$. Hence,  if  there exists  a portfolio $\bm{w}_0 \in \mathcal{W}^+$ satisfying 
$ (\bm{w}_0^\top\bm{\mu} - t)_- >\lambda$, then  
  $\E^{G}[(\bm{w}_0^\top\bm{X} - t)_-] >\lambda$ for all $G\in \mathcal M$,  which means that  the set $ \mathcal{M}_{\bm{w}_0 , \lambda} $  
   is empty.  
  Thus,   $\sup_{G \in   \mathcal{M}_{\bm{w}_0,   \lambda}}\E^{G}[(\bm{w}_0^\top\bm{X} - t)^2_+]= \sup \emptyset=-\infty$, which implies 
        $$\min_{\bm{w} \in \mathcal{W}^+} \ \sup_{G \in   \mathcal{M}_{\bm{w},  \lambda}} ~~ \E^{G}[(\bm{w}^\top\bm{X} - t)^2_+]   =-\infty.$$ 
   Therefore,     any portfolio $\bm{w}$ satisfying  $(\bm{w}^\top\bm{\mu} - t)_- >\lambda$  is a solution to problem \eqref{pro:pf-m-EL<lambda}. Note that for  any portfolio $\bm{w}$ satisfying  $(\bm{w}^\top\bm{\mu} - t)_- >\lambda$, 
    it holds that   $ \E^{G}[(\bm{w}^\top\bm{X} - t)_-] \ge  (\bm{w}^\top\bm{\mu} - t)_- > \lambda$, which 
     means that  the expected excess profits with  such solutions or portfolios  exceed the desirable level $\lambda$. In this sense, such solutions or portfolios       are acceptable and reasonable.  However, to exclude such trivial solutions,  in this subsection,  we assume that  for any 
    $\bm{w} \in \mathcal{W}^+$,   it holds that $ (\bm{w}^\top\bm{\mu} - t)_-  \le \lambda$,
   which is   equivalent to assume that 
\begin{equation}
\label{con-w}
\sup_{\bm{w} \in \mathcal{W}^+} (\bm{w}^\top\bm{\mu} - t)_- \leq \lambda. 
\end{equation} 
Note that $\mathcal{W}^+$ only contains non-negative  allocations added up to 1 and $(x-t)_-$ is decreasing in $x$.  Hence, condition \eqref{con-w} is further  equivalent to
\begin{equation} \label{eq:0526-1}
\big (\min_{1 \leq i \leq d} \, \mu_i - t\big )_- \leq \lambda.
\end{equation}
In this subsection, we assume that condition  \eqref{eq:0526-1} holds. 
  
  To solve problem \eqref{pro:pf-m-EL<lambda} with  $\mathcal{M} = 
 \mathcal{M}(\bm{\mu}, \bm{\Sigma})$ or  $\mathcal{M}_S(\bm{\mu}, \bm{\Sigma})$ in \eqref{Mwl},
 we first  use   arguments similar to those in  the proof  of Lemma \ref{lem:multi-S}    
  to  show (the proofs are omitted)   that 
 \begin{equation*}
 \label{sup-Lw,m=Lm}
 \sup_{G \in  \mathcal{M}_{\bm{w},  \lambda}}\, \E^{G}[(\bm{w}^{\top}\bm{X} - t)^2_+] 
 =\sup_{F \in \mathcal{L}_{\bm{w},  \lambda}}\, \E^{F}[(X - t)^2_+],
 \end{equation*}
where  
  $\mathcal{L}_{\bm{w},  \lambda} =  \mathcal{L}_{\lambda}\big (\bm{w^{\top}\mu},\, \sqrt{ \bm{w^{\top}}{\bf \Sigma}\bm{w}}\, \big  )$ for   $\mathcal{M} = 
 \mathcal{M}(\bm{\mu}, \bm{\Sigma})$ in \eqref{Mwl}; 
   $\mathcal{L}_{\bm{w},  \lambda} = \mathcal{L}_{S, \lambda}\left (\bm{w^{\top}\mu},\, \sqrt{ \bm{w^{\top}}{\bf \Sigma}\bm{w}}\, \right )$
  for   $\mathcal{M}   =\mathcal{M}_S(\bm{\mu}, \bm{\Sigma})$ in \eqref{Mwl}; and $X$ is a random variable with 
 a distribution belonging to  $\mathcal{L}_{\bm{w},  \lambda}$. Then, 
  we 
  immediately obtain the following solutions  to  problem \eqref{pro:pf-m-EL<lambda}:
 
      \begin{proposition}
\label{prop:upml} 
Suppose $t \in \mathbb{R}$, $\lambda >0$ and ${\bm \mu}\in \R^d$ satisfy  \eqref{eq:0526-1}, and the covariance matrix $\mathbf{\Sigma}$ is positive definite. 
Then, the optimal portfolio $\bm{w}^*_{\lambda}$ for problem \eqref  {pro:pf-m-EL<lambda}  with ${\cal M} = \mathcal{M}(\bm{\mu}, {\bf \Sigma})$ is solved by 
  \begin{equation}
 \bm{w}^*_{\lambda} \ = \ \argmin_{\bm{w} \in \mathcal{W}^+} ~h_{\lambda, t} \big ( \bm{w^{\top}\mu}, \sqrt{\bm{w^{\top}}{\bf \Sigma}\bm{w}} \,  \big),
  \end{equation}
and the optimal portfolio $\bm{w}^*_{S, \lambda}$ for problem  \eqref  {pro:pf-m-EL<lambda}   with ${\cal M} = \mathcal{M}_S(\bm{\mu}, {\bf \Sigma})$ is solved by 
  \begin{equation}
 \bm{w}^*_{S, \lambda} \ = \ \argmin_{\bm{w} \in \mathcal{W}^+} ~h_{S,\lambda, t}\big  (\bm{w^{\top}\mu}, \sqrt{\bm{w^{\top}}{\bf \Sigma}\bm{w}}\, \big ),
  \end{equation}
  where 
  $$h_{\lambda,t}(\mu, \sigma)
= \sup_{F \in \mathcal{L}_\lambda(\mu, \sigma)}\E^F[(X-t)^2_+]~~~{\rm and}~~~h_{S,\lambda,t}(\mu, \sigma)
= \sup_{F \in \mathcal{L}_{S, \lambda}(\mu, \sigma)}\E^F[(X-t)^2_+],$$
which have  the explicit expressions  presented  in  Theorem \ref{th:constraint} and 
Theorem \ref{th-2nd-TVS-S-m},  respectively.

        \end{proposition}

\begin{remark} In problem  \eqref{pro:pf-m-EL<lambda}, we restrict  portfolios to be in 
 $\mathcal{W}^+ = \{ \bm{w} \in \R^d: \bm{w}^\top\bm{e} = 1, \, \bm{w} \geq  0 \}$, which indicates   that an investor does not 
 short sell any stock or denotes  the rule that   short-selling is not allowed. We point out that this restriction is necessary for  problem  \eqref{pro:pf-m-EL<lambda} to have  non-trivial  solutions. If fact, if $\mathcal{W}^+$ in problem \eqref{pro:pf-m-EL<lambda} 
 is relaxed to $\mathcal{W} = \{ \bm{w} \in \R^d: \bm{w}^\top\bm{e} = 1  \}$ as used in problem  \eqref{pro:pf-m-EL-g}, then for any $\lambda >0$, 
  there always exists  a portfolio $\bm{w}_0 \in \mathcal{W}^+$ satisfying 
$ (\bm{w}_0^\top\bm{\mu} - t)_- >\lambda$ or  
  $\E^{G}[(\bm{w}_0^\top\bm{X} - t)_-]  \ge   (\bm{w}_0^\top\bm{\mu} - t)_-  >\lambda$ for all $G\in \mathcal M$,  which means that  the set $ \mathcal{M}_{\bm{w}_0,  \lambda} $  
  is empty. 
  Thus,   $\sup_{G \in   \mathcal{M}_{\bm{w}_0,   \lambda}} ~~ \E^{G}[(\bm{w}_0^\top\bm{X} - t)^2_+]= \sup \emptyset=-\infty$, which implies 
     $$\min_{\bm{w} \in \mathcal{W}^+} \ \sup_{G \in   \mathcal{M}_{\bm{w},  \lambda}} ~~ \E^{G}[(\bm{w}^\top\bm{X} - t)^2_+]   =-\infty.$$ 
   Therefore,     the portfolio $\bm{w}_0$  is a solution to problem \eqref{pro:pf-m-EL<lambda}. \hfill $\Box$
\end{remark}

\section{Numerical experiments    with  real  financial data}
\label{sec:num}

In this section,  we conduct a numerical study using real financial data to calculate the optimal portfolios derived in Section \ref{sec:ER-TSV} and compare the investment performances of the optimal portfolios   with several existing portfolio selection methods related to the   models proposed in this paper. For this study, we select 12 stocks from the four largest sectors (Technology, Health Care, Financials, and Consumer Discretionary) of the S\&P 500, choosing three with the highest market capitalizations from each sector.\footnote{Selected stocks are AAPL, MSFT, GOOG, JPM, BAC, BRK-B, PFE, JNJ, UNH, HD, TSLA, AMZN}
We use data from a four-year period starting from January 1, 2019, to January 1, 2023, which include  1008 observations of daily stock prices. The daily losses are expressed by percentage and calculated by $l_t = -(V_{t+1} - V_t)/ V_t$, where $V_t$ is the close price on trading day $t$. Note that the positive value represents the loss and negative value represents the gain.

We aim to compare investment performance across several existing  models related to the  models proposed  in this paper, including:
\begin{itemize}
    \item[(a)] TSV model:  Minimizing the  target semi-variance of the portfolio loss and formulated as (see e.g., \cite{CHZ11})
     \begin{equation}
\label{CHZ11}
      \min_{\bm{w} \in \mathcal{W}} \ \sup_{G \in   \mathcal{M}(\bm{\mu}, {\bf \Sigma})}\, \E^{G}[(\bm{w}^\top\bm{X} - t)^2_+].
\end{equation} 

\item[(b)] M-TSV-S model:  Minimizing the target semi-variance of the portfolio loss, incorporating
the symmetric information of loss vectors and the constraint on the expected portfolio loss, and  proposed in \eqref{pro:pf-m-EL-g} and 
solved  in Proposition \ref{prop:upm-S}.

\item[(c)] EEP-TSV model:  Minimizing the target semi-variance of the portfolio loss over undesirable scenarios,  and  proposed in \eqref{pro:pf-m-EL<lambda}  and 
solved  in Proposition \ref{prop:upml}.

\item[(d)] EEP-TSV-S model:   Minimizing the target semi-variance of the portfolio loss over undesirable scenarios,   incorporating
the symmetric information of loss vectors, and  proposed in \eqref{pro:pf-m-EL<lambda}  and 
solved  in Proposition \ref{prop:upml}.

\item[(e)] M-V model: The classical mean-variance model formulated  in \eqref{C-MV-model}.

\end{itemize}

We construct portfolio rebalancing strategies by using  the optimal solutions  to
 models (a)-(e)  listed above. The experiment is set up as follows. 
 The initial portfolio $\bm{w}_0^{*}$ is calculated on January 3, 2020 using the data from January 2, 2019 to January 2, 2020 as 
the  in-sample dataset   (252 trading days).  We compute the out-of-sample portfolio returns as $-\bm{w}_0^{* \top}\bm{\hat{l}}_0$, where $\bm{\hat{l}}_0$ represents the daily loss on January 3, 2021. We proceed to optimize the portfolio selections on a daily basis using a rolling window approach and subsequently rebalance the portfolio. This involves using the preceding 755 trading days to calculate the optimal portfolio  $\bm{w}_t^{*}$ for trading day $t$, serving 
as an updated
 portfolio  for each trading day starting from January 3, 2020. The resulting portfolio returns  $-\bm{w}_t^{* \top}\bm{\hat{l}}_t$  for trading day $t$  are obtained using the out-of-sample return vector 
 $\bm{\hat{l}}_t$ and the rebalanced portfolio weights  $\bm{w}_t^{*}$. In the TSV-based models 
 (a) and (b), the parameters $u$, $v_0$, $v_1$, and $v_2$ defined in \eqref{v} are also updated as the data rolls forward. These parameters rely on sample mean and sample covariance, which  evolve with rebalance process over time. To conduct the numerical experiment, parameters need to be chosen for models (a)-(e). We give  the following general guidelines for selecting parameters: the target return $-t$ in models (a) and (b), the desirable minimum level $\lambda$ for the expected excess profit over the target return in models (c) and (d), the maximum expected loss level $\nu$ in models (b) and (e).
     \begin{enumerate}
 \item[(1)]  Note that in this paper, positive (negative) values of a portfolio loss random variable $X$ represent losses (returns). Initially, it might seem logical for an investor to choose a higher target return $-t$ or a lower threshold loss level $t$ to expect better investment performance. However,  the expected excess profit $ \E[(X-t)_-]$ increases with $t$, while the expected downside risk $ \E[(X-t)_+]$ decreases with $t$. Thus, opting for a lower threshold loss level $t$ results in higher expected downside risk. Consequently, a reasonable choice for the target return $-t$ is to set it slightly larger than the sample mean of the daily returns of the selected stocks. Equivalently, the threshold loss level $t$ can be set slightly smaller than the sample mean of the daily losses of the selected stocks. 
      
 \item[(2)]  The target return $-t$ represents an investor's goal. The investor is satisfied if the expected excess profit $\E[(X-t)_-]$ is positive. Note that $\E[(X-t)_+] = \E[(X-t)_-] + \E[X] - t$. This implies that a higher desirable minimum level $\lambda$ for the expected excess profit will result in a higher lower bound for the expected downside risk. Specifically, $ \E[(X-t)_-] > \lambda$ is equivalent to $ \E[(X-t)_+] > \lambda + \E[X] - t$. Therefore, a reasonable choice for the desirable minimum level $\lambda$ for the expected excess profit is a small positive value. 
 
\item[(3)] Note that $\E[X] \le \nu$ is equivalent to $\E[-X] \ge -\nu$; where $\E[-X]$ represents the expected return, and $-t$ is the target return. Thus, it is natural to require $-t \ge -\nu$ or equivalently $\nu \ge t$. Additionally, a high value of $\nu$ is not desirable. Therefore, a reasonable choice for the maximum expected loss level $\nu$ is to set it slightly larger than  $t$. 
    \end{enumerate}
     
 According to the above guidelines, in this experiment,   we choose  
a  target return  $-t = 0.003$ for all the TSV-based models (a)-(d); a desirable   minimum  level 
 $\lambda = 0.015$ for  the expected excess profit in    models  (c) and (d);  a maximum   expected  loss level $\nu = -0.001$ for  models (b) and (e).

\begin{figure}[t]
\centering
\includegraphics[width=15cm]{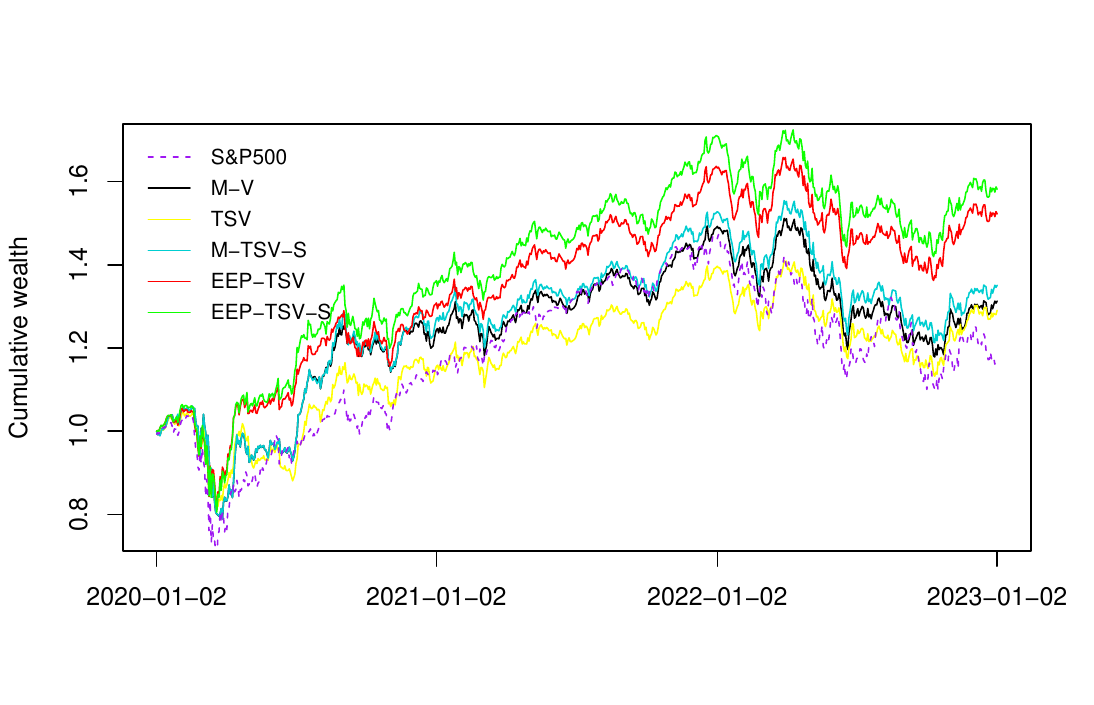}
\caption{Cumulative wealth comparison across portfolio rebalancing strategies based on models (a) to (e) and 
the S\&P 500 index. The target return   $t = -0.003$ for all the TSV-based models; desirable  level $\lambda = 0.015$ for the expected excess profit in   all the EEP-TSV-based  models; risk level $\nu = -0.001$ for the M-V and M-TSV-S models. 
}\label{fig:wealth1}
\end{figure}

Figure \ref{fig:wealth1} shows the cumulative wealth of a portfolio comprised of the 12 selected stocks   under the strategies derived from models (a)-(e) listed above.  We also include 
 the performance of  the S\&P 500 index in Figure \ref{fig:wealth1} to compare it with the performance of these portfolio selection models  (a)-(e).  
 It is evident that all the strategies, except for the TSV model, outperform the passive investment strategy of the S\&P 500 index.
 We can also see from Figure \ref{fig:wealth1} that the EEP-TSV-S model (d) outperforms all the other models listed above.
The expected excess profit constraint enhances the capability of controlling the downside risk. Additionally, the additional information regarding the symmetry of the loss distribution
 (as indicated Figure \ref{fig:sym} for several stocks), greatly improves the practicality of the models proposed in this paper.  Other models, including the M-TSV-S model, which incorporates only symmetric information, and the EEP-TSV model, which incorporates only the expected excess profit constraint, also perform well in our experiment. Therefore, incorporating both symmetric information and expected downside constraints into portfolio selection models can significantly improve investment performance when using the models proposed in this paper.

\section{Concluding remarks}
\label{sec:end}

In this paper, we explore the worst-case target semi-variance of a random loss within mean-variance uncertainty sets, considering additional distributional information such as  symmetry and non-negativity of the random loss. We introduce new robust portfolio selection models wherein investors aim to minimize the worst-case target semi-variances of a portfolio loss over undesirable scenarios while incorporating the symmetric information of the loss vectors of the portfolio. The contributions of the paper are threefold:  Firstly, it complements the study of \cite{CHZ11}, where the worst-case target semi-variance was derived for an arbitrary random  loss. We extend this by deriving the worst-case target semi-variances for symmetric or non-negative losses, thus obtain the  results for the worst-case target semi-variances    corresponding to the worst-case expected regrets investigated in  \cite{J77}. Secondly, we derive the worst-case target semi-variances of an arbitrary, symmetric, or non-negative random loss over undesirable scenarios, which represent the main concerns of an investor or decision maker. These worst-case values provide insights into the greatest deviation in downside risk among these adverse conditions. Thirdly, based on the worst-case target semi-variances derived in this paper, we propose new robust portfolio selection models where investors minimize the worst-case target semi-variances over undesirable uncertainty sets. These proposed models emphasize controlling downside risk by minimizing the worst-case value of the second moment of the downside risk of a portfolio while limiting the first moment. As illustrated in numerical experiments, the investment performance with the optimal strategies derived from the proposed models outperforms the classical mean-variance strategy and several other existing models. We believe the results and models developed in this paper have more potential and will explore more applications in future research.
 
\section*{Acknowledgment}
J. Cai  acknowledges financial support from the Natural Sciences and Engineering Research Council of Canada (RGPIN-2022-03354).   T. Mao acknowledges financial support from the National Natural Science Foundation of China (Grant 12371476).

\section{Appendix}
\label{sec:App}

\subsection{Proofs of results in Section \ref{WC-SD}}
\noindent{\bf Proof of {Lemma} \ref{lem:3}.} Denote  $\underline{F} =  \essinf \, F  $ and $\overline{F} = \esssup \, F$. We show the result by considering the following two cases. 

{\bf Case (i)}:   Assume  that  $\underline{F},  \, \overline{F} \in\R$.
 For any $F\in \mathcal{L} (\mu, \sigma)$,  there exists $p =1-q \in (0,1)$ such that
 $p \, \underline{F}  +q \,  \overline{F} =\mu$ as $\mu \in (\underline{F},  \, \overline{F})$. Let
  $G=    [\underline{F}, p; \overline{F},q] $ that is a two-point distribution.
 Clearly,   $ \E^G[X]= p \, \underline{F}  + q \,  \overline{F} = \mu$. In addition,  note that the number of the sign changes of $F-G$ is one.
  By  Theorem 3.A.44 of \cite{SS07},
  we have that  $F\le_{\rm cx } G $\footnote{For two distributions $F$ and $G$, we say that $F\le_{\rm cx } G$ if $\E^F[u(X)] \le \E^G[u(Y)]$ for all convex functions $u$, where $X$ and $Y$ are two random variables that follow  $F$ and $G$, respectively.} and thus, ${\rm Var}^G(X)\ge \sigma^2$.  For any
  $\epsilon \ge0$, define the two-point distribution $G_\epsilon$ as  $G_\epsilon =  [\underline{F}  +\epsilon q,\, p;~ \overline{F} - \epsilon p,\,q].
 $
We have
 $\E^{G_\epsilon} [X]=  p \, (\underline{F} +\epsilon q) +q \,  (\overline{F} - \epsilon p)  = \mu$ and
 \begin{align*}
 {\rm Var}^{G_\epsilon} (X) &= p \, (\underline{F} +\epsilon q)^2 +q \,  (\overline{F} - \epsilon p)^2 -   \mu^2
 = pq \epsilon^2 - 2pq(\overline{F}     - \underline{F})\epsilon + p\underline{F}^2+q \overline{F}^2 -\mu^2.
    \end{align*}
Thus,  ${\rm Var}^{G_\epsilon} (X)$  is  a  quadratic function of $\epsilon$ with ${\rm Var}^{G_0} (X) =  p\underline{F}^2+q \overline{F}^2 -\mu^2 =  {\rm Var}^G(X)  \ge \sigma^2$ and   ${\rm Var}^{G_{\epsilon_0}} (X)=0$, where
   $\epsilon_0 = \overline{F}     - \underline{F}.$
 Since   ${\rm Var}^{G_\epsilon} (X)$ is   continuous and
   decreasing in $\epsilon \in [0, \epsilon_0]$,  there must exist $\epsilon_\sigma  \in [0,\epsilon_0)$ such that  ${\rm Var}^{G_{\epsilon_\sigma}} (X)=\sigma^2$. Therefore, $G_{\epsilon_\sigma} \in \mathcal{L}_2(\mu, \sigma)$ and
 the support of $G_{\epsilon_\delta}$ belongs to $[\underline{F},   \, \overline{F}]$ as
 $\underline{F}\le \underline{F}  +\epsilon_\sigma \, q  \le \underline{F}  +\epsilon_0 q =  \underline{F}  + (\overline{F}     - \underline{F})q < \overline{F} $
 and 
$ \overline{F} \ge   \overline{F} - \epsilon_\sigma \, p \ge  \overline{F} - \epsilon_0 \, p  = \overline{F} -(\overline{F}     - \underline{F})   \, p   >  \underline{F}. $

{\bf Case (ii)}:   Assume   that $\underline{F} = -\infty$ or $\overline{F} =\infty$. In this case,  it suffices to show that we can find a distribution $F_0\in \mathcal L(\mu,\sigma)$ with  bounded  support satisfying $[\underline{F}_0, \, \overline{F}_0] \subseteq [\underline{F}, \, \overline{F}]$.  To see it, we only give the proof  of the   case
 that $\underline{F} \in \R$ and $\overline{F} =\infty$ as
the other cases can be proved  easily by using the similar  arguments  for the case
 that $\underline{F} \in \R$ and $\overline{F} =\infty$.
  For the case $\underline{F} \in \R$ and $\overline{F} =\infty$,   note that $\mu>\underline{F}$. For $\alpha\in (0,1)$, define two-point random variable $X_\alpha$ with the following probability function:
$$
  \p\Big (X_\alpha = \mu -\sigma \sqrt{\frac{1-\alpha}{\alpha}} \, \Big )  =\alpha =1-\p \Big (X_\alpha = \mu +\sigma \sqrt{\frac{\alpha}{1-\alpha}} \, \Big ).
$$
We have $\E[X_\alpha]=\mu$ and ${\rm Var}(X_\alpha) = \sigma^2$ for any $\alpha\in (0,1)$.
There exists $\alpha_0\in (0,1)$ such that $ \mu -\sigma \sqrt{\frac{1-\alpha_0}{\alpha_0}} >  \underline{F}$ as  $\mu>\underline{F}$.
 Thus,
the distribution of $X_{\alpha_0}$ is the desired distribution $F_0$. Therefore,  we  complete the first part of the proof of Lemma \ref{lem:3}.

We next consider the case that  $F$ is symmetric. For any $F\in \mathcal{L}_S (\mu, \sigma)$, if $\underline{F}=\essinf \, F \in \R$ and $ \overline{F}= \esssup \, F\in\R$,
it holds that $\overline{F}-\mu=\mu-\underline{F}$. In this case,
in the above proof of case (i), it holds that $p=q=1/2$ and  the two-point distribution  $G_\epsilon =  [\underline{F}  +\epsilon q,\, p;~ \overline{F} - \epsilon p,\,q]$ is symmetric  about $\mu$. Hence, there exists a two-point symmetric distribution  $F^*\in \mathcal{L}_{2, S}(\mu, \sigma)$ such that   the support of $F^*$ belongs to $[\essinf \, F,\, \esssup \, F]$. Moreover,  If $\mu=0$, then
$\overline{F}=-\underline{F}$ and $F^*= 0.5 \, \delta_x + 0.5 \, \delta_{-x}$ for some $x\in (0, \,  \esssup \, F]$, where $\delta_x$ means 
a degenerate distribution at $x$. This completes  the proof of Lemma \ref{lem:3}.
\qed

 \noindent{\it \bf Proof of Theorem \ref{th-2nd-TVS-S}.}
According to Remark \ref{re-mu=0}, we assume $\mu=0$ in the following proof. We  consider the three cases that  (i) $t \ge 0$; (ii)  $t\le -\sigma$;  and (iii) $ -\sigma < t < 0$ below.

{\bf Case (i)}:   Assume  $t \ge 0$. In this case, it must hold that $0\le  \p(X>t)=\E^F[1_{\{X> t\}}] <1$ for any $F\in  \mathcal{L}_{S}(0, \sigma)$.
If $\p(X>t)=\E^F[1_{\{X> t\}}] =0$ for an  $F\in  \mathcal{L}_{S}(0, \sigma)$,   then $\E^{F}[(X-t)_{+}^2]=0$ for this  $F$. Hence,
to determine  $ \sup_{F \in \mathcal{L}_S(0, \sigma)} \E^{F}[(X-t)_{+}^2]$, we only need to consider
 those distributions $F$ in  $\mathcal{L}_{S}(0, \sigma)$ satisfying   $0 < \p(X>t) =  \E^F[1_{\{X> t\}}] < 1$. For  such a distribution
  $F$ in  $\mathcal{L}_{S}(0, \sigma)$ satisfying   $0 < \p(X>t) =  \E^F[1_{\{X> t\}}] < 1$,
      let $p=\p(X>t)=\p(X<-t) \in (0,1/2]$, and let $G_t$ be the  distribution of
the (conditional)  random variable $[X|X>t]$, then
 \begin{equation*}
   G_t(x) =\p(X \le x | X >t) =
        \begin{cases}
  0,   & x \le t, \\
 \frac{F(x)-F(t)}{1-F(t) }, & x >t.
    \end{cases}
\end{equation*}
Note that $\essinf G_t \ge t $ and $\esssup G_t = \esssup F$. Denote the mean and variance of $G_t$ by $\mu_t$ and $\sigma_t^2$, we have
$\mu_t=\E[X|X>t]$ and $\mu_t^2+\sigma_t^2=\E[X^2|X>t]$.
  By applying Lemma \ref{lem:3} to the distribution $G_t$, we know that
there exists a two-point distribution $F_t \in  \mathcal{L}_2(\mu_t, \sigma_t)$ such that $\E^{F_t}[X_t]=\mu_t$,
  $\E^{F_t}[X_t^2]= \mu_t^2+\sigma_t^2$,  and the support of $F_t$ belongs to $[\essinf G_t, \, \esssup G_t ] \subset [t, \esssup F]$, where $X_t\sim F_t$.
 Note that $X_t \ge  t \ge 0$.

    Denote the probability function of $X_t$ by $[x_{1,t}, p_{1,t}; \,x_{2,t}, p_{2,t} ]$, where $0< p_{i, t} < 1$, $i=1,2$, and $p_{1, t}+p_{2,t} =1$.
      For any
  $\epsilon\ge0$, define a  random variable $X^*_\epsilon$,
    with  distribution $F_\epsilon^*$, as
    \begin{equation}
    \label{X*-1}
      X^*_\epsilon = (X_t+\epsilon)1_{\{U>1-p\}} + 0\, 1_{\{p\le U\le 1-p\}} - (X_t+\epsilon)1_{\{U<p\}},
   \end{equation}
    where $U\sim {\rm U}[0,1]$ is a uniform random variable  independent of $X_t$.  Thus,   $ X^*_\epsilon$ is a   five-point  random variable   valued on $\{-x_{2,t}-\epsilon, \, -x_{1,t}-\epsilon,  \, 0, \,
  x_{1,t}+\epsilon, \, x_{2,t}+\epsilon \}$.  For any $x \ge 0$, it holds that
   \begin{align*}
  \p(X^*_\epsilon >x ) & = \p((X_t+\epsilon)1_{\{U>1-p\}}>x) =  \p( X_t+\epsilon>x,\, U>1-p) = p \, \p(X_t+\epsilon>x)
\\  & =  p \, \p(-X_t-\epsilon<-x) =  \p(-(X_t+\epsilon)1_{\{U<p\}}<-x) =\p(X^*_\epsilon <-x ) ,
 \end{align*}
where the third equality  follows from the independence of $X_t$ and $U$. Similarly,  for any $x<0$,  it holds that
  \begin{align*}
  \p(X^*_\epsilon >x ) & = \p((U > 1-p) \cup (p \le U \le 1-p) \cup ( -(X_t+\epsilon)1_{\{U<p\}}>x)) \\
  & =p + 1-2p +  \p( -(X_t+\epsilon)1_{\{U<p\}}>x) \\
  & =1-p +  \p( X_t+\epsilon< -x,\, U<p) = 1-p+p \, \p(X_t+\epsilon<-x)
\\
&=  \p((U <p) \cup (p \le U \le 1-p) \cup ((X_t+\epsilon)1_{\{U>1-p\}}<-x)) =\p(X^*_\epsilon <-x ).
 \end{align*}
 Therefore $X^*_\epsilon $ is a symmetric random variable at $0$ and $\E[X^*_\epsilon]=0$.  Moreover, note that $X_t \ge t \ge 0$. Thus,     for any $\epsilon\ge0$,   we have
     \begin{align*} \E[(X^*_\epsilon-t)_{+}^2]& =  \E[\((X_t+ \epsilon )1_{\{ U>1-p\}} -t\)^2]  =  \E[(X_t+ \epsilon -t)^2 1_{\{ U>1-p\}}]
  =\E[(X_t+ \epsilon -t)^2 ] \, \p(U>1-p ) \\
    & \ge p \,  \E[(X_t -t)^2 ] 
     =   \p(X>t) \left (\E[X_t^2] -2t \E[X_t] +t^2     \right)\\
    & =  \p(X>t)\left (\E[X^2|X>t] -2t \E[X|X>t] +t^2     \right) \\
 & = \p(X > t)\,\E^F[(X-t)^2|X > t] = \E^{F}[(X-t)_{+}^2], 
     \end{align*}
where the inequality follows from $X_t+ \epsilon -t\ge X_t-t \ge  0$. In addition,   \begin{align*}{\rm Var}(X^*_0) &= \E[(X^*_0)^2]  = \E[X_t^2 \, 1_{\{U>1-p\}} + X_t^2 \, 1_{\{U<p\}}] = 2p \, \E[X_t^2]= 2p \, \E[X^2|X>t] \\
& =2\E[X^21_{\{X>t\}}]  = \E[X^21_{\{X>t\} \cup \{X<-t\}} ] \le\E[X^2] =\sigma^2,
\end{align*}
where the second equality follows from the independence between $X_t$ and $U$, the forth equality follows from $p=\p(X>t)$, and the fifth equality follows from that $X$ is symmetric  at $0$.

 Clearly, ${\rm Var}(X^*_\epsilon)  =
\E[(X_t+\epsilon)^2 \, 1_{\{X> t\}} + (X_t+\epsilon)^2 \, 1_{\{X <  -t\}}]$ is a quadratic function of   $\epsilon$ with
$ {\rm Var}(X^*_\epsilon) \to \infty $ as $\epsilon \to \infty$. There exists $\epsilon_\delta \ge 0$ such that ${\rm Var}(X^*_{\epsilon_\delta})  =
\sigma^2$. Hence,
the distribution of $X^*_{\epsilon_\delta}$ belongs to $\mathcal{L}_{S}(0, \sigma)$ and $\E^{F}[(X-t)_{+}^2] \le \E^{F}[(X^*_{\epsilon_\delta}-t)_{+}^2]$, where $X^*_{\epsilon_\delta}$ has a five-point symmetric distribution about 0.
 Therefore, for $t>0$, $ \sup_{F \in \mathcal{L}_{S}(0, \sigma)}\E^{F}[(X-t)_{+}^2]=  \sup_{F \in \mathcal{L}_{5,S}(0, \sigma)}\E^{F}[(X-t)_{+}^2]$.
 Note that the probability function of   the five-point symmetric random variable $X^*_{\epsilon_\delta}$ has the expression
 $[-x_2, p_2;  \, -x_1, p_1; \,  0,p_0;  \, x_1, p_1;  \, x_2, p_2]$, where $0\le t \le x_1 \le x_2$,  $0 < p_1+p_2 \le 1/2$, $0 \le p_1 < 1/2$,   $0 \le  p_2<1/2$, and $0 \le p_0 <1$. Thus,
 the problem $ \sup_{F \in \mathcal{L}_{S}(0, \sigma)}\E^{F}[(X-t)_{+}^2]$ is equivalent to the problem
    \begin{align}
     \label{S=5S-t>0}
         \sup_{(p_1, p_2, x_1, x_2) \in [0, \frac{1}{2}]^2  \times \R_+^2}  &~~ \, \, p_1(x_1-t)^2+p_2(x_2-t)^2, \\
    {\rm s.t.}&~~  \, \, t\le x_1\le x_2,   \, \,  \ 0< p_1+p_2\le 1/2,
     \, \,  \ p_1 \, x_1^2+p_2 \, x_2^2=\sigma^2/2.\nonumber
    \end{align}
    One can verify that the supremum   of
  problem \eqref{S=5S-t>0}  is equal to  $ \sigma^2/2.$ To see it, note that  for any feasible solution of the problem \eqref{S=5S-t>0}, it holds that
   \begin{align*}
   p_1(x_1-t)^2+p_2(x_2-t)^2
   & =  p_1x_1^2+p_2x_2^2 + (p_1+p_2) t^2 -2t(p_1x_1+p_2x_2)\\
   & = \frac{\sigma^2} 2 + t \left[(p_1+p_2) t  -2(p_1x_1+p_2x_2)\right]
   \, \le \,  \frac{\sigma^2} 2 ,
   \end{align*}
  where the inequality follows from that   $(p_1+p_2) t  -2(p_1x_1+p_2x_2)\le 0$ as $x_1,x_2\ge t$.
   On the other hand, for $\epsilon >0$ small enough, take
   $
   p_1=0, ~~ p_2= \epsilon,~~ x_1 = t,~~x_2 =  \sqrt{\frac{\sigma^2}{2\epsilon}} .
   $
   We have the objective function in \eqref{S=5S-t>0}  is
    \begin{align*}
    p_1(x_1-t)^2+p_2(x_2-t)^2 &= \epsilon  \Big ( \sqrt{\frac{\sigma^2}{2\epsilon}} -t \Big )^2 
    =   \Big ( \sqrt{\frac{\sigma^2 }{2}}- \sqrt{\epsilon} t \Big )^2  \,  \to  \, \frac{\sigma^2} 2~~{\rm as}~\epsilon\to0.
    \end{align*}
     Note that for any $F\in \mathcal{L}_{S}(0, \sigma)$, $$\E^F[(X-t)_+^2] = \frac{\E^F[(X-t)_+^2]  + \E^F[(X+t)_-^2] }2 < \frac{\E[X^2] }2 =\frac{\sigma^2}2,$$
where the first equality follows from the symmetry of $F$ at $0$, and the inequality follows from $\E[X^2] = (\E[X_+^2] +\E[X_-^2] )/2$ and $\E^F[(X-t)_+^2]  + \E^F[(X+t)_-^2]$ is strictly decreasing in $t\ge 0$.
   The supremum  $\sigma^2/2$   of  problem \eqref{S=5S-t>0}  is the limit of 
 $ \E^{F\epsilon}[(X-t)_{+}^2]$ as $\epsilon \to 0$, where  $F\epsilon$ 
 is  the following  three-point symmetric distribution:  
    $\big [\mu-\sqrt{\frac{\sigma^2}{2\epsilon}}, \, \epsilon; ~\mu, \, 1-2\epsilon; ~\mu+\sqrt{\frac{\sigma^2}{2\epsilon}},
    \, \epsilon\big ].
    $
    
  {\bf Case (ii)}:   For $t\le -\sigma$, on one hand, note that for any $F\in \mathcal{L}_{S}(0, \sigma)$, we have
    $
      \E^F[(X-t)^2_+] \le  \E^F[(X-t)^2_+]+ \E^F[(X-t)^2_-]= \E^F[(X-t)^2] = \sigma^2+t^2 .
    $
    On the other hand, take $X\sim F$ as
    $
     \p(X=-\sigma) =\p(X=\sigma) =1/2.
    $
    We have $X\ge t$ a.s., and  $
      \E[(X-t)^2_+] =\frac12 (-\sigma -t)^2 +\frac12 (\sigma-t)^2= \sigma^2+t^2 .
    $
Therefore, we have $\sup_{F  \in \mathcal{L}_{S}(0, \sigma)} \E^F[(X-t)_+^2]=\sigma^2+t^2.$ 

{\bf Case (iii)}: For $-\sigma < t< 0$,  we first show that for any   $F\in  \mathcal{L}_{S}(0, \sigma)$, there exists a six-point distribution $G\in \mathcal{L}_{S}(0, \sigma)$ such that $\E^F[(X-t)_+^2]\le \E^G[(X-t)_+^2]$.
Note that in the case that $-\sigma < t< 0$, we must have $ \p(t \le X\le -t) < 1$ as otherwise $ \|X\|\le -t $ a.s. which yields a contradiction with $\sigma>- t$.
We then have   $ p:=\p(X>-t)>0$, and $  \p(X< t)=p>0$ by symmetry.
Applying  Lemma \ref{lem:3} to the distribution of  $[X|X>-t]$, we see that
there exists a two-point distribution $F_t$ with support on $[-t,\infty),$ such that $\E^{F_t}[X_t]=\E^F[X|X> -t]$ and $\E^{F_t}[X_t^2]=\E^F[X^2|X> -t]$, where $X_t\sim F_t$ and $X_t \ge -t >0$.

If $p<1/2$, then $ \p(t \le X \le -t )>0$, and  by applying Lemma \ref{lem:3} to $[X | t \le X \le -t ]$, we see that   there exists $x\in (0,-t]$ such that  $\E^{G_x} [Y_t^2] = \E[X^2 |t\le  X\le -t]$,  where $G_x =\delta_x/2+\delta_{-x}/2$ and $Y_t \sim G_x$.
   Define
    \begin{equation}
    \label{X*-2}
      X^*_\epsilon = (X_t+\epsilon)1_{\{U>1-p\}} + x1_{\{1/2< U\le 1-p\}} - x1_{\{p\le U\le 1/2\}} - (X_t+\epsilon)1_{\{U<p\}},
    \end{equation}
    where $x\in (0,-t]$, $\epsilon\ge0$, and $U\sim {\rm U}[0,1]$ is independent from $X_t$ and $p=\p(X>-t) \in (0,1/2]$.
   Otherwise, if $p=1/2$, we still employ the definition of the random variable $X^*_\epsilon$  by \eqref{X*-2}, which reduces to \begin{equation*}
      X^*_\epsilon = (X_t+\epsilon)1_{\{U>1-p\}}  - (X_t+\epsilon)1_{\{U<p\}}.
    \end{equation*}
    In both cases,   $ X^*_\epsilon$ is a   six-point  random variable   valued on $\{-x_{2,t}-\epsilon, \, -x_{1,t}-\epsilon,  \, -x, \, x,
  x_{1,t}+\epsilon, \, x_{2,t}+\epsilon \}$, where $x \in (0,-t]$ and  $-t \le  x_{1, t} < x_{2,t} $.

Similar to Case (i), we can verify that   the distribution of $ X^*_\epsilon $ is symmetric about 0 and
 that $\E [X^*_\epsilon ]=0$ and $  \E[(X^*_\epsilon-t)_{+}^2]\ge \E^{F}[(X-t)_{+}^2]$ for any $\epsilon\ge0$, and ${\rm Var}[(X^*_0)]\le \sigma^2$ for $\epsilon=0$. Moreover, one can verify that ${\rm Var}[(X^*_\epsilon)]$ is a quadratic function of    $\epsilon\ge0$. There exists $\epsilon_\delta \ge 0$ such that the distribution of $X^*_{\epsilon_\delta}$ belongs to $\mathcal{L}_{S}(0, \sigma)$.  Denote the set $\mathcal{L}^*_{6,S}(0, \sigma)$ by
 \begin{align*}
 \mathcal{L}^*_{6,S}(0, \sigma)=\{&[-x_3, p_3;  \, -x_2, p_2; \,  -x_1,p_1;  \,  x_1, p_1; x_2, p_2;  \, x_3, p_3]: \\
 &0<x_1 \le -t \le  x_2  < x_3,  \ p_1+p_2 +p_3 = 1/2, \  0 \le p_i \le 1/2, \ \mbox{for}\  i=1,2,3 \}.
 \end{align*}
 Then, $\mathcal{L}^*_{6,S}(0, \sigma) \subset \mathcal{L}_{6,S}(0, \sigma)$ and
  the distribution of $X^*_{\epsilon_\delta}$ belongs to $\mathcal{L}^*_{6,S}(0, \sigma)$. Hence, it holds that 
  $ \sup_{F \in \mathcal{L}_{S}(0, \sigma)}\E^{F}[(X-t)_{+}^2]=  \sup_{F \in \mathcal{L}^*_{6,S}(0, \sigma)}\E^{F}[(X-t)_{+}^2]$.
     Note that $\E^F[(X-t)_+^2] + \E[(X-t)_-^2] =\sigma^2 +t^2$ is fixed for any $F\in \mathcal L(0,\sigma)$.
     Thus, we have \begin{align}\label{eq-0719-1} \sup_{F \in \mathcal{L}^*_{6,S}(0, \sigma)}\E^{F}[(X-t)_{+}^2]= \sigma^2 +t^2 - \inf_{F \in \mathcal{L}^*_{6,S}(0, \sigma)}\E^{F}[(X-t)_{-}^2].\end{align}
     Note that  the problem $\inf_{F \in \mathcal{L}^*_{6,S}(0, \sigma)}\E^{F}[(X-t)_{-}^2]$  is equivalent to the problem
         \begin{align}
    \inf_{  (p_1,p_2, p_3,x_1, x_2, x_3) \in [0, 1/2]^3  \times \R_+^3} & ~~ \, \,  p_2(x_2+t)^2+p_3(x_3+t)^2,  \label{inf-S-t<0}\\
    {\rm s.t.}&~~ 0<  x_1\le  -t\le x_2\le x_3, \, \,p_1+ p_2+p_3=1/2, \nonumber
    \\
    &~~ \, \,p_1x_1^2+ p_2x_2^2+p_3x_3^2=\sigma^2/2, \nonumber
    \end{align}
 as $(-x)_-=(x)_+$,  $x_2+t \ge 0$, and $x_3+t\ge0$.

 By $-t< \sigma$, we know that the constraints in problem \eqref{inf-S-t<0} can not be satisfied  at $x_2=x_3=-t$.  Note that for any feasible solution of the problem \eqref{inf-S-t<0}, $(p_1,p_2, p_3,x_1, x_2, x_3)$, if $x_1<-t$, then take $\delta\in (0,-t-x_1)$ and
    $
     (p_1,p_2, p_3,x_1+\delta, x_2-\delta_1, x_3-\delta_2),
    $
    where $\delta_1,\delta_2\ge 0$ satisfy  $-t\le x_2-\delta_1\le x_3-\delta_2$ and $p_1(x_1+\delta)^2+ p_2(x_2-\delta_1)^2+p_3(x_3-\delta_2)^2=\sigma^2/2$. It holds that the value of the objective function  at the new feasible solution $(p_1,p_2, p_3,x_1+\delta, x_2-\delta_1, x_3-\delta_2)$ is strictly smaller than that at
    $(p_1,p_2, p_3,x_1, x_2, x_3)$.  Therefore,  the  infimum of problem \eqref{inf-S-t<0} is attainable at $x_1=-t$, which   implies that   problem \eqref{inf-S-t<0} is equivalent to
  \begin{align}
    \label{eq:617-1}
    \min_{  (p_1,p_2, p_3, x_2, x_3) \in [0, 1/2]^3  \times \R_+^2} & ~~ \, \,  p_2(x_2+t)^2+p_3(x_3+t)^2, \nonumber \\
    {\rm s.t.} &~~  -t\le x_2\le x_3, \, \, \, p_1+ p_2+p_3=1/2,
    \, \, \, p_1t^2+ p_2x_2^2+p_3x_3^2=\sigma^2/2.
    \end{align}
One can verify that for any feasible solution of \eqref{eq:617-1}, it holds that
 \begin{align*}
  p_2(x_2+t)^2+p_3(x_3+t)^2
  & = p_1 (x_1+t)^2 + p_2(x_2+t)^2+p_3(x_3+t)^2 ~~~~~~~~~(x_1=-t)\\
 & = p_1x_1^2+p_2 x_2^2  +p_3 x_3^2  + (p_1+p_2+p_3)t^2+ 2t(p_1x_1+p_2x_2+p_3x_3) \nonumber\\
 & = \frac {\sigma^2} 2 + \frac {t^2} 2 + 2t(p_1x_1+p_2x_2+p_3x_3).
   \end{align*}
    Noting that  $t<0$,  the problem \eqref{eq:617-1} is equivalent to
     \begin{align} \label{eq-0716-1}
    \max_{  (p_1,p_2, p_3, x_2, x_3) \in [0, 1/2]^3  \times \R_+^2} & ~~ \, \,  2p_1x_1+2p_2x_2+2p_3x_3, \nonumber \\
    {\rm s.t.}&~~  -t\le x_2\le x_3, \ \ 2p_1+ 2p_2+2p_3=1, \ \ 
    2p_1t^2+ 2p_2x_2^2+2p_3x_3^2=\sigma^2,
    \end{align}
For any feasible solution, we have
$
2p_1x_1+2p_2x_2+2p_3x_3 \le  \sqrt{2p_1t^2+ 2p_2x_2^2+2p_3x_3^2} = \sigma.
$
On the other hand, take
$
p_1=0,~ p_2=0,~ x_1=x_2=-t,~p_3=1/2,~ x_3=\sigma>-t.
$
We have $2p_1x_1+2p_2x_2+2p_3x_3 =\sigma. $
Therefore, we have the  supremum  of  problem \eqref{eq-0716-1} is $\sigma$, and thus, the infimum  of problem \eqref{eq:617-1} 
is $\frac {\sigma^2} 2 + \frac {t^2} 2  + t \sigma = \frac {(\sigma+t)^2}{2}.$ It thus follows from \eqref{eq-0719-1}  that 
is
$ \sup_{F \in \mathcal{L}_{S}(\mu, \sigma)}\E^{F}[(X-t)_{+}^2] =\sigma^2 +t^2-\frac {(\sigma+t)^2}{2} = \frac {(\sigma -t)^2}{2}. 
$ This completes the  proof. 
  \qed

\subsection{Proofs of results in Section \ref{sec:WC-ER}}
\noindent{\it \bf Proof of Theorem \ref{th:constraint}.}  We show the result by considering the following two cases. 

{\bf Case 1}:  If  $\lambda =(\mu-t)_-$, then 
  by $\lambda>0$, we have  $ \mu<t$.  In this case, the constraint $\E^F[(X-t)_-] \le \lambda$ is 
   $\E^{F}[(X-t)_-] = (\E^{F}[X]-t)_-$, that is, the Jensen inequality reduces to equality. This implies  $X\le t$ a.s. Also, by  $t>\mu$, we have $ \mathcal{L}_{\lambda}(\mu, \sigma)$ is not an empty set.  Then $\E^{F}[(X-t)_+^2]=0$ for any $F\in \mathcal{L}_{\lambda}(\mu, \sigma)$.

 {\bf Case 2}:  If $\lambda >(\mu-t)_-$,  define $m=\mu-t+\lambda$. Noting  that  $m\ge \lambda -(\mu-t)_-> 0$ by the assumption, and $m>\mu-t$ by $\lambda>0$, we have $m > (\mu-t)_+.$
Note  that for any $F \in \mathcal{L}_\lambda(\mu, \sigma)$, it holds that $\E^F[(X-t)_+] =\E^F[(X-t)_-] +\mu-t$, and thus, $\E^F[(X-t)_-]\le \lambda$ if and only if $\E^F[(X-t)_+]\le m$.
Therefore,  problem \eqref{pro:budget-l} is equivalent to  the following optimization problem:
\begin{equation}
\label{pro:budget-1}
\begin{aligned}
&
\sup_{F \in \mathcal{L}_{m}^*(\mu, \sigma)} \E^{F}[(X-t)^2_+],
  \end{aligned}
 \end{equation}
where $
    \mathcal{L}_{m}^*(\mu, \sigma) = \big \{  F \in \mathcal{L}(\mu, \sigma):   \    \E^F[(X-t)_+]\leq m  \big \}.  $
We next solve problem \eqref{pro:budget-1}.  
 Note that $(x-t)_- \ge 0 $ and it is convex in $x \in \R$. In addition, $x^2$ is convex in $x \in \R$  and non-decreasing  in $x \ge 0$. Hence
$(x-t)^2_- $ is convex in $x \in \R$. Thus, for any $F \in \mathcal{L}_{m}^*(\mu, \sigma)$, by Jensen's inequality, we have
$\E^{F}[(X-t)_{-}^2]    \ge  (\mu-t)^2_-  $ and $
\E^{F}[(X-t)_{-}]    \ge (\mu-t)_-,$
which imply
\begin{align}
\label{inf>t+}
   \inf_{F \in \mathcal{L}_{m}^*(\mu, \sigma)}    \E^{F}[(X-t)_{-}^2]   & \ge  (\mu-t)^2_-, \ \ \ \   \ \ \ \  \inf_{F \in \mathcal{L}_{m}^*(\mu, \sigma)}    \E^{F}[(X-t)_{-}]    \ge  (\mu-t)_-.
\end{align}
For $\delta\in (0,1)$, let $F_\delta$ be a two-point distribution of the  random variable  $X_\delta$ that is defined   as
 \begin{equation} \label{X-delta}
 \p(X_\delta=\mu-\epsilon) = 1-\delta,~~~\p(X_\delta=\mu+ M) =\delta,
 \end{equation}
 where  $\epsilon>0$,   $M>0$  satisfy  $\E[X_\delta]=\mu$  and ${\rm Var}(X_\delta) =\sigma^2$, which imply
  $ (\mu- \epsilon) (1-\delta) +(\mu+M) \delta =\mu$ 
  and 
  $(\mu-\epsilon)^2 (1-\delta) +(\mu+M)^2\delta =\mu^2+\sigma^2.$ Solving the  two equations, we have
\begin{equation}
      M = \sigma \sqrt{\frac{1-\delta}{\delta}}, \ \ \ \ \  \epsilon = \sigma \sqrt{\frac{\delta}{1-\delta}}. \label{M-ep}
      \end{equation}
      Note that $F_\delta \in \mathcal{L}(\mu, \sigma)$ for any $\delta \in (0, 1)$ and $\E^{F_\delta}[(X_\delta-t)_-^2 ] =(1-\delta) \,   (\mu -\epsilon  -t)_-^2  + \delta \,   (\mu+M-t)_-^2$
      and  $\E^{F_\delta}[(X_\delta-t)_- ] =(1-\delta) \,   ( \mu-\epsilon  -t)_-  + \delta \,   (\mu+M-t)_-.$ When $\delta$ is sufficiently  small, say $\delta \to 0$, we see from \eqref{M-ep} that  $\epsilon \to 0$,   $M \to \infty$,
 $\mu+M-t>0$,  and
\begin{align}
\E^{F_\delta}[(X_\delta-t)_-^2 ] &=  (1-\delta)  ( \mu-\epsilon  -t)_-^2   \to (\mu-t)_-^2  \ \ \ \mbox{as} \ \,  \delta \to 0, 
 \label{E-limit}
 \end{align}
 where the limit holds since the functions  $(x-t)_-^2$ is  continuous in $x \in \R$. Thus, we have   as $\delta \to 0$
\begin{align}
\E^{F_\delta}[(X_\delta-t)_+] = \E^{F_\delta}[X_\delta-t] + \E^{F_\delta}[(X_\delta-t)_-] & \to  \mu - t +(\mu-t)_-    = 
(\mu-t)_+ < m . \label{limit<m}
 \end{align}
 By \eqref{E-limit} and \eqref{limit<m}, we see that 
 $\E^{F_\delta}[(X_\delta-t)_+]   \nearrow (\mu-t)_+ < m$   as  $ \delta \searrow   0$. 
 Hence,   there exists a series $\{\delta_n:  \delta_n \in (0, 1),~ n=1,2,...\}$ such that  $\E^{F_{\delta_n}}[(X_{\delta_n}-t)_+]<m$ or
  $F_{\delta_n} \in  \mathcal{L}_{m}^*(\mu, \sigma)$, 
  which, together with \eqref{inf>t+} implies 
    $
    \inf_{F \in \mathcal{L}_{m}^*(\mu, \sigma)} \E^{F}[(X-t)_{-}^2]=(\mu-t)_-^2 $ 
    and 
    $\inf_{F \in \mathcal{L}_{m}^*(\mu, \sigma)} \E^{F}[(X-t)_{-}]=(\mu-t)_-.
  $
 Since $(x)^2_+=x^2-(x)^2_-$, we have
     $
    \sup_{F \in \mathcal{L}_{m}^*(\mu, \sigma)} \E^{F}[(X-t)_{+}^2]
    = \sigma^2 + (\mu-t)^2  \, - \,  \inf_{F \in \mathcal{L}_{m}^*(\mu, \sigma)} \E^{F}[(X-t)^2_-] 
    =   \sigma^2 +  (\mu-t)^2 \, - (\mu-t)_-^2=    \sigma^2 +   (\mu-t)_+^2.
    $
 which yields \eqref{eq:constraint} for this case.   
   \qed

\noindent{\bf Proof of Proposition \ref{Prop:1}.}   It is equivalent to show that  the set $\mathcal{L}^+_{\lambda}(\mu, \sigma)$ is empty if and only if 
\begin{align}\label{eq:nonemty-1}
\lambda = (\mu-t)_-~~{\rm and}~~\sigma^2 >  \mu (t-\mu).
 \end{align} 
First we show the ``if" part.  Suppose that \eqref{eq:nonemty-1} holds and  $ \mathcal{L}^+_{\lambda}(\mu, \sigma)$ is not empty.  Take $F\in \mathcal{L}^+_{\lambda}(\mu, \sigma)$ and let $X\sim F$. We have 
 $  (\mu-t)_-\le \E^F[(X-t)_-]\le\lambda = (\mu-t)_-,$
 where the first inequality follows from the Jensen inequality, and the second one follows from the constraint of the set $\mathcal{L}^+_{\lambda}(\mu, \sigma)$. Therefore, both the inequalities should be equality, that is, $\E^F[(X-t)_-] = (\mu-t)_-=t-\mu$, where the second equality follows from $ (\mu-t)_-=\lambda >0$ and thus, $t>\mu$.  This is equivalent to $ \E^F[(X-t)_+ ]=\E^F[(X-t) ] + \E^F[(X-t)_-] = 0$,
    and thus, $X\le t$ a.s.  Define a two-point distribution  $F^* = \big [0, 1- \frac{\mu}{t}; \,  t, \frac{\mu}{t} \big ]$ and $X^*\sim F^*$.
     Since $\E^F[X]=\mu>0$ and $0\le X \le t$ a.s., 
     we have  $X \le_{\rm cx} X^*$,  and thus, 
    $ 
 \sigma^2= {\rm Var}^F(X) \le   {\rm Var}^{F^*}(X^*) =\mu(t-\mu).
      $
    This yields a contradiction to \eqref{eq:nonemty-1} and thus completes the proof of the ``if" part.

    We next consider the ``only if" part. Suppose that  \eqref{eq:nonemty-1} does not hold. We have following two cases. 
    \begin{itemize}
    \item [(i)] If $\lambda > (\mu-t)_-$,    then  $\sup_{F \in \mathcal{L}^+_{\lambda}(\mu, \sigma)} \E^{F}[(X-t)_{+}^2]$ 
    is finite, which is proved in  Corollary \ref{coro:positive<m}, 
     and thus, the   set $   \mathcal{L}^+_{\lambda}(\mu, \sigma)  $ is not empty.
    \item [(ii)] If $\lambda = (\mu-t)_-$ and $\sigma^2 \le  \mu (t-\mu)$ then for $\epsilon \in [0, t]$, define 
   the two-point distribution  $F_\epsilon = [p\epsilon, 1-p; \,  t-(1-p)\epsilon, p]$ and $X_\epsilon\sim F_\epsilon$,  where $p=\mu/t$. We have $\E^{F_\epsilon}[X_\epsilon]=\mu$, $0\le X_\epsilon \le t$, and 
    $
    {\rm Var}^{F_\epsilon} (X_\epsilon) 
    $
   satisfies that $ {\rm Var}^{F_\epsilon} (X_\epsilon) $ is continuous in $\epsilon$,  
   $
    {\rm Var}^{F_0} (X_0)  =\mu(t-\mu)$  and ${\rm Var}^{F_t} (X_t)=0.  
   $
   There must exist $\epsilon$ such that $ {\rm Var}^{F_\epsilon} (X_\epsilon) =\sigma^2$.
Therefore, the set is not empty.    \end{itemize}
  Combining the above two cases, we complete the proof.  
 \qed

\noindent{\it \bf Proof of Corollary \ref{coro:positive<m}.}   
{\bf Case 1}: For the case $\lambda =(\mu-t)_-$,  we have $ \mu<t$. The constraint $\E^F[(X-t)_-]\le\lambda$ reduces to $\E^F[(X-t)_-]=t-\mu$, that is, $X\le t$ a.s..
Also, by Proposition \ref{Prop:1}, we have $ \mathcal{L}^+_{\lambda}(\mu, \sigma) $ is not empty and thus $ \sup_{F \in \mathcal{L}^+_{\lambda}(\mu, \sigma)} \E^{F}[(X-t)_{+}^2] =0.$

 {\bf Case 2}: For the case $\lambda >(\mu-t)_-$,  
first note that the problem $ \sup_{F \in \mathcal{L}^+_{\lambda}(\mu, \sigma)} \E^{F}[(X-t)_{+}^2] $  is bounded from above by the problem \eqref{eq:constraint}.   Also, 
let $F_\delta$ be the distribution of the two-point distribution
$X_\delta$ defined in  \eqref{X-delta}. From the proof of  Theorem \ref{th:constraint}, we see that $\mu -\epsilon >0$ as $\delta \to 0$. Hence
$F_\delta \in \mathcal{L}^+_{\lambda}(\mu, \sigma)$ as $\delta \to 0$.   
By
$ \sup_{F \in \mathcal{L}^+_{\lambda}(\mu, \sigma)} \E^{F}[(X-t)_{+}^2]=
              \sigma^2 + (\mu-t)_+^2$, which, together with  Theorem \ref{th:constraint}, implies that     Corollary \ref{coro:positive<m} holds.
\qed

 \noindent{\bf Proof of Proposition \ref{Prop:2}.} 
 The proof is similar to that of Proposition \ref{Prop:1}  and the only difference is that the two-point distribution $F^*$  in the proof of Proposition \ref{Prop:1}   is defined as $\big [t, \frac{1}{2}; \,  2\mu-t, \frac{1}{2} \big ]$,  whose variance is $ (t-\mu)^2$. The details of  the proof are omitted. 
   \qed

\noindent{\it \bf Proof of Lemma \ref{eq:0919-1}.}  In this case, problem \eqref{pro:budget-l} is equivalent to  the following optimization problem:
\begin{equation}
\label{pro:budget<m-sym}
\begin{aligned}
&
\sup_{F \in \mathcal{L}^*_{S, m}(\mu, \sigma)} \E^{F}[(X-t)^2_+],
  \end{aligned}
 \end{equation}
where  $m=\lambda+\mu-t$, and
$
  \mathcal{L}_{S,m}^*(\mu, \sigma) = \big \{ F \in \mathcal{L}_S(\mu, \sigma):      \E^F[(X-t)_+]
  \leq m   \}. $
 According to Remark \ref{re-mu=0},  
we  assume $\mu=0$ in the following proof.  It suffices to show for any symmetric   distribution $F\in \mathcal{L}_S(\mu, \sigma)$, there exists 
a six-point symmetric   distribution   $F^*\in \mathcal{L}_{6,S}(\mu, \sigma)$ satisfying 
\begin{equation}
 \label{EF<EF*} 
 \E^{F^*}[(X-t)_+] \le m, \  \  \ \ \   
\E^{F}[(X-t)^2_+] \le \E^{F^*}[(X-t)^2_+].
 \end{equation} 
We next consider the following  three cases. 

{\bf Case (i)}: 
If $\p( X>|t|)\in (0,1/2)$,   by Lemma \ref{lem:3}, there exist a two-point distribution  $F_t$ and  a symmetric distribution $G$ such that the support of $F_t$ belongs to $(|t|,\infty)$;   the support of $G$ with support belongs to $[-|t|,|t|]$; 
 $\E^{F_t} [X_t] =\E^F[X|X>|t|]~ {\rm and}  ~\E^{F_t} [X_t^2] = \E^F[X^2 |X>|t|];
$
and
$ \E^{G} [Y] =\E^F[X|-|t|\le X \le|t|]~ {\rm and}  ~\E^{G} [Y] = \E^F[X^2 | -|t|\le X \le |t|],
$
where $X_t\sim F_t$ and $Y\sim G.$
Define $F^* = p \tilde{F}_t + (1-2p) G+ p F_t$,  where $p=\p(X>|t|)$ and $\tilde{F}_t$ is the distribution of $-X_t$. 
It is easy to  verify that $F^*$ is a six-point symmetric distribution or  $F^* \in \mathcal L_{S}(\mu,\sigma) $ and 
satisfy 
$\E^{F^*} [(X-t)_+] = \E^{F} [(X-t)_+]$ and 
$ \E^{F^*} [(X-t)_+^2] =\E^{F} [(X-t)_+^2]$. 
Therefore, this  six-point symmetric distribution    $F^*$ satisfies \eqref{EF<EF*}.
 
 {\bf Case (ii)}:   If $\p( X>|t|)= \frac{1}{2}$,  then the distribution $F^* = \frac{1}{2} \, \tilde{F}_t +  \frac{1}{2} \,  F_t$   satisfies \eqref{EF<EF*}.  

{\bf Case (iii)}: 
 If $\p( X>|t|)=0$,  the distribution $F^* =G$   satisfies \eqref{EF<EF*}.
   Thus, by combining the above three cases, we complete the proof.
\qed

\noindent{\it \bf Proof of Theorem \ref{th-2nd-TVS-S-m}. }  We show the result by considering the following three cases:

    {\bf Case (a)}. Assume that   $\sigma \leq m$. 
     First consider the subcase  $t>\mu$.  Note that the supremum of  $\sup_{F \in \mathcal{L}_{S}(\mu, \sigma)}\E^{F}[(X-t)_{+}^2]$ is an upper bound of the  supremum of   problem  \eqref{pro:budgetapp<m-sym-k} whose  supremum    is $\sigma^2/2$. The  supremum $\sigma^2/2$  is the limit of 
 $ \E^{F\epsilon}[(X-t)_{+}^2]$ as $\epsilon \to 0$, and $F_\epsilon$ is also a feasible distribution of the problem   \eqref{pro:budgetapp<m-sym-k} for $\epsilon>0$ small enough. 
$F\epsilon$  is  the following  three-point symmetric distribution:  
    $\big [\mu-\sqrt{\frac{\sigma^2}{2\epsilon}}, \, \epsilon; ~\mu, \, 1-2\epsilon; ~\mu+\sqrt{\frac{\sigma^2}{2\epsilon}},
    \, \epsilon \big ].
    $
  For the subcase $t\le \mu$, note that the worst-case distribution of the problem $\sup_{F \in \mathcal{L}_{S}(\mu, \sigma)}\E^{F}[(X-t)_{+}^2]$ is $[\mu-\sigma, \, 0.5;~\mu+\sigma,\, 0.5]$ which is also a feasible distribution of the problem \eqref{pro:budgetapp<m-sym-k}. Therefore, if $\mu -m \le t\le \mu-\sigma$, the supremum of problem  \eqref{pro:budgetapp<m-sym-k} is $ \sigma^2 + (t-\mu)^2$; If $ \mu - \sigma <t \le \mu $, the supremum   of problem \eqref{pro:budgetapp<m-sym-k} is $\frac{1}{2}(\mu- t+\sigma)^2$.
  
    {\bf Case (b).} Assume $ m < \sigma \le 2m$. For this case, we consider the following three sub-cases:
     \begin{itemize}
\item [(i)]
  If $t >\mu$, the supremum   of  problem  \eqref{pro:budgetapp<m-sym-k}
  is $ {\sigma^2}/2$ that     is the limit of 
 $ \E^{F\epsilon}[(X-t)_{+}^2]$ as $\epsilon \to 0$, where  $F\epsilon$ 
 is  the following  three-point symmetric distribution:  
    $\big [\mu-\sqrt{\frac{\sigma^2}{2\epsilon}}, \, \epsilon; ~\mu, \, 1-2\epsilon; ~\mu+\sqrt{\frac{\sigma^2}{2\epsilon}},
    \, \epsilon\big ].
    $

 \item [(ii)] If $ \mu+\sigma -2m<t\le \mu $, the     supremum  of  problem  \eqref{pro:budgetapp<m-sym-k}
  is $\frac{1}{2}(\mu- t+\sigma)^2$, and one worst-case distribution  is $
 [\mu-\sigma, \, 0.5;~\mu+\sigma,\, 0.5].
$ 

     \item [(iii)] If $\mu - m <t\le \mu+\sigma -2m$, the   supremum     of  problem \eqref{pro:budgetapp<m-sym-k}
  is $  \frac{1}{2}\sigma^2 + 2m(\mu - t) - \frac{(t-\mu)^2}{2}$, and one worst-case distribution is $$
 [\mu+t-(m+t-\mu)/p,\, p;~\mu+t, 0.5-p;~\mu-t, \, 0.5-p;~ \mu-t+(m+t-\mu)/p,\, p] 
$$ with $p=2(m+t-\mu)^2/(\sigma^2+3(t-\mu)^2+4m(t-\mu))\in [0, \frac{1}{2}]$.
 \end{itemize}
The sub-cases (i) and (ii) follow the same arguments in Case (a). It remains to show 
 (iii). According to Remark \ref{re-mu=0},  we assume $\mu=0$ in the following proof. By Lemma  \ref{eq:0919-1} and its proof, we have  problem  \eqref{pro:budget<m-sym-m1}  is also equivalent to 
 \begin{equation}
\label{pro:budget<m-sym-sixpoints}
\begin{aligned}
&
\sup_{F \in \mathcal{L}^*_{6,S, m}(\mu, \sigma)} \E^{F}[(X-t)^2_+],
  \end{aligned}
 \end{equation}
 where $\mathcal{L}_{6, S,m}^*(\mu, \sigma) = \big \{F \in \mathcal{L}_{S,m}^*(\mu, \sigma): F \ \mbox{is a six-point distribution}\}. $ Therefore, we have that the problem   \eqref{pro:budgetapp<m-sym-k} is equivalent to the following problem  
       \begin{align}\label{eq:0919-3} 
    \max_{  (p_1,p_2, p_3, x_1,x_2, x_3) \in [0, 1/2]^3  \times \R_+^3} & ~~ \, \,    p_1(\tilde{x}_1-t)^2+p_1(x_1-t)^2+ p_2(x_2-t)^2+ p_3(x_3-t)^2, \nonumber \\
    {\rm s.t.}&~~t \le \tilde{x}_1\le 0\le x_1\le s
\le x_2\le x_3,
    \\
    &~~   \, \,2p_1+ 2p_2+2p_3=1, \, \, \ \sum_{i=1}^3 p_i(\tilde{x}_i^2+x_i^2)=\sigma^2, \nonumber\\
     &~~ \,  \, p_1(\tilde{x}_1-t)+p_1(x_1-t)+ p_2(x_2-t)+p_3(x_3-t)\le m\nonumber
    \end{align}
   for $t<0$, in the sense of that if the  maximizer of  problem 
   \eqref{eq:0919-3}  is  $(p_1^*,p_2^*, p_3^*, x_1^*,x_2^*, x_3^*) $, then the worst-case  distribution is $F^*=[\tilde{x}_3^*,p_3^*; \tilde{x}_2^*,p_2^*; \tilde{x}_1^*,p_1^*; x_1^*,p_1^*; x_2^*, p_3^*; x_3^*,p_3^*]$, where $\tilde{x}_i^*=-x_i^*$, $i=1,2,3$, and $s=-t$. 
  Note that for any feasible solution $(p_1,p_2, p_3, x_1,x_2, x_3)$, it holds that
$$
p_3(\tilde{x}_3-t)^2+ p_2(\tilde{x}_2-t)^2  +  p_1(\tilde{x}_1-t)^2+p_1(x_1-t)^2+ p_2(x_2-t)^2+ p_3(x_3-t)^2 =  \sigma^2 + t^2 
$$
which  is a constant independent from the decision variables.  Here $\tilde{x}_i=-x_i$, $i=1,2,3$. 
Therefore, we have that the problem  \eqref{eq:0919-3} is equivalent to 
  \begin{align}\label{eq:0919-5} 
    \min_{  (p_1,p_2, p_3, x_1,x_2, x_3) \in [0, 1/2]^3  \times \R^3} & ~~ \, \,   p_3(\tilde{x}_3-t)^2+ p_2(\tilde{x}_2-t)^2 , \nonumber \\
    {\rm s.t.}&~~ t \le \tilde{x}_1\le 0\le x_1\le s
\le x_2\le x_3,
    \\
    &~~   \, \,2p_1+ 2p_2+2p_3=1, \, \, \  \sum_{i=1}^3 p_i(\tilde{x}_i^2+x_i^2)=\sigma^2, \nonumber\\  
     &~~ \,  \, p_1(\tilde{x}_1-t)+p_1(x_1-t)+ p_2(x_2-t)+p_3(x_3-t)\le m. \nonumber
    \end{align}
Noting that $p_3(\tilde{x}_3-t)+ p_2(\tilde{x}_2-t)  + p_1(\tilde{x}_1-t)+p_1(x_1-t)+ p_2(x_2-t)+p_3(x_3-t)=-t$  and taking $s=-t\ge 0$, we have $\tilde{x}_i-t =-x_i-t = s-{x}_i$, $i=1,2,3$ and that the last constraint is equivalent to $p_3(\tilde{x}_3-t)+p_2(\tilde{x}_2-t)\ge  -t-m$, that is,  $p_3(s-{x}_3)+p_2(s-{x}_2)\ge  -t-m$.
Thus, 
the above problem \eqref{eq:0919-5}  is again equivalent to
  \begin{align}\label{eq:0919-6} 
    \min_{  (p_1,p_2, p_3, x_1,x_2, x_3) \in [0, 1/2]^3  \times \R^3} & ~~ \, \,   p_2( {x}_2-s)^2 +p_3(x_3-s)^2 , \nonumber \\
    {\rm s.t.}&~~ \mu \le x_1\le s
\le x_2\le x_3, \, \, \ 2p_1+ 2p_2+2p_3=1,
    \\
    &~~   \, \,2\sum_{i=1}^3p_i x_i^2=\sigma^2, 
    \,  \,  \ p_2(x_2-s)+p_3(x_3-s)\le m+t.\nonumber
    \end{align}
If  any feasible solution $(p_1,p_2, p_3, x_1,x_2, x_3) $ satisfies $p_i>0$, $i=1,2,3$, $x_1<s$ and $s\le x_2<x_3$, then for $\epsilon>0$ small enough, take $
x_1^*=x_1+\epsilon_1,~~x_2^*=x_2+\epsilon p_3,~~x_3^*=x_3-\epsilon p_2,
$
where $\epsilon_1$ is taken such that $2\sum_{i=1}^3p_i(x_i^*)^2=\sigma^2$. One can verify that $(p_1,p_2, p_3, x_1^*,x_2^*, x_3^*) $ is still a feasible solution for $\epsilon>0$ small enough and 
 the objective value becomes smaller. Therefore, we have the worst-case distribution is  either a two-point distribution or satisfies $x_1=s$.
For $x_1=s$, the objective value of the problem \eqref{eq:0919-6}  is 
$
p_2(x_2-s)^2+p_3(x_3-s)^2 
 = \frac{\sigma^2}2 +\frac{s^2}{2} = \frac{s^2+\sigma^2}2 .
 $
 Therefore the     minimal value  of the problem \eqref{eq:0919-6}  is the minimal value of $\frac{s^2+\sigma^2}2$ and the optimal value to the problem  
 \begin{align}\label{eq:0919-7} 
    \min_{  (p_1,p_2, x_1,x_2 ) \in [0, 1/2]^2  \times \R^2} & ~~ \, \,   p_1(x_1-s)_+^2+ p_2(x_2-s)_+^2 , \nonumber \\
    {\rm s.t.}&~~ 0\le x_1
\le x_2, \, \, \ 2p_1+ 2p_2=1,
    \\
    &~~  2p_1x_1^2+ 2p_2x_2^2=\sigma^2,
      \,  \,  \ p_1(x_1-s)_++p_2(x_2-s)_+\le m-s.\nonumber
    \end{align}
  Noting that if $x_1\ge s$, then
 we have the objective function of the problem \eqref{eq:0919-7}  also equals to $\frac{s^2+\sigma^2}2$.   Therefore,   the problem \eqref{eq:0919-6} is equivalent to the problem \eqref{eq:0919-7}.

Note that for any feasible solution of \eqref{eq:0919-7}, it holds that 
$
p_2(x_2-s)^2 =\frac{s^2+\sigma^2}2 - p_1(x_1-s)^2  \le \frac{s^2+\sigma^2}2.
$
So,  the problem \eqref{eq:0919-3} is equivalent to the following problem  \begin{align}\label{eq:0919-7-1} 
    \min_{  (p_1,p_2, x_1,x_2 ) \in [0, 1/2]^2  \times \R^2} & ~~ \, \,  p_2(x_2-s)^2 , \nonumber \\
    {\rm s.t.}&~~ 0\le x_1\le s
\le x_2, \, \, \ 2p_1+ 2p_2=1, \\
    &~~   \, \,2p_1x_1^2+ 2p_2x_2^2=\sigma^2, 
      \,  \, \   p_2(x_2-s)\le m-s.\nonumber
    \end{align}
   Taking $x=x_1$, $z=x_2-s$, and  $p=2p_2=1-2p_1$, the above problem is again equivalent to
  \begin{align}\label{eq:1005-1} 
    \min_{  (p , x,z ) \in [0, 1]  \times \R_+^2} & ~~ \, \, \frac12 pz^2 , \nonumber \\
    {\rm s.t.}&~~ x\le s,  \, \, \  pz\le 2(m-s),
     \, \, \ (1-p)x^2+ pz^2+ps^2+2pzs=\sigma^2.
     \end{align} 
    One can verify that for any feasible solution of the problem \eqref{eq:1005-1}, it holds that 
    \begin{align*}
  pz^2  & = \sigma^2 -(1-p)x^2-ps^2-2pzs \\ 
        & \ge \sigma^2 - s^2-2pzs \ge \sigma^2 - s^2-4(m-s)s =\sigma^2 +3 s^2-4ms,
    \end{align*}
    where the  equality comes from the last constraint of the problem \eqref{eq:1005-1},   the first inequality comes from   $x\le s$ and the second inequality follows from that $pz\le 2(m-s)$.
    On the other hand, taking $x=s$, $pz=2(m-s)$, we have  
 $ pz^2 =\sigma^2 +3 s^2-4ms $.  Therefore the supremum   of the problem \eqref{eq:0919-7-1}  is $\frac {\sigma^2  +3 s^2-4ms} 2$,  and the  supremum   of   problem \eqref{eq:0919-3} is
 $  \sigma^2 + s^2 - \frac{\sigma^2 +3 s^2-4ms} 2= \frac{\sigma^2 - s^2} 2+2ms.
 $
 We recover the general case by letting $s = \mu - t$, then the  supremum  of problem \eqref{pro:budgetapp<m-sym-k} is 
$ \frac{\sigma^2 - (\mu - t)^2}{2} + 2 m (\mu - t).$

{\bf Case (c)}: Assume  $\sigma > 2m$.     This case follows the same arguments in Case (a), but with a modification for the case $t\le \mu$. Specifically, if $ \mu - m <t \le \mu $, the  supremum  of problem \eqref{pro:budgetapp<m-sym-k} is $\frac{1}{2}(\mu- t+\sigma)^2$.

 Combining the above three cases, we complete the proof. 
\qed

\end{document}